%% file: main.tex
\documentclass[lettersize,journal]{IEEEtran}
\usepackage{colortbl}
\usepackage[table]{xcolor}

\usepackage{tikz}
\usepackage{algorithm}
\usepackage{algorithmicx}
\usepackage{algpseudocode}
\usepackage{url}
\usepackage{float} 
\usepackage{algpseudocode} 

\usepackage{amsmath,amsfonts,amssymb,mathrsfs,amsthm,amsbsy,graphicx,color,bm,algpseudocode,balance,mathtools,xfrac,nccmath,siunitx,cite,xspace}
\usepackage{multirow}

\usepackage[font=scriptsize,labelfont=bf]{caption}

\newcolumntype{C}[1]{>{\centering\let\newline\\\arraybackslash\hspace{0pt}}m{#1}}

\usepackage{pgfplots}
\pgfplotsset{compat=newest}
\usetikzlibrary{plotmarks}
\usetikzlibrary{arrows.meta}
\usepgfplotslibrary{patchplots}
\usepackage{grffile}
\usetikzlibrary{plotmarks}

 \usepackage[utf8]{inputenc}
\usepackage{booktabs, caption, makecell}

\usepackage{threeparttable}

\newcommand{\bc}{\text{BackCom}\xspace}

 \theoremstyle{definition}
 \makeatletter
\newcommand{\printfnsymbol}[1]{%
  \textsuperscript{\@fnsymbol{#1}}%
}
\makeatother

\newcommand{\qa}{{\bf a}}
\newcommand{\qb}{{\bf b}}

\newcommand{\qf}{{\bf f}}
\newcommand{\qg}{{\bf g}}

\newcommand{\qr}{{\bf r}}

\newcommand{\qv}{{\bf v}}
\newcommand{\qw}{{\bf w}}

\newcommand{\qy}{{\bf y}}

\newcommand{\qI}{{\bf I}}

\newcommand{\qQ}{{\bf Q}}

\newcommand{\ettall}{\emph{et al.}}

\newcommand{\SINR}{\mathtt{SINR}}
\newcommand{\Ex}{\mathbb{E}}
\newcommand{\Sn}{\sigma_n^2}
\newcommand{\MR}{\mathsf{MR}}

\newcommand{\dl}{\mathtt{dl}}
\newcommand{\ul}{\mathtt{ul}}

\newcommand{\alphml}{\alpha_{ml}}

\DeclareMathOperator{\MM}{\mathcal{M}}
\DeclareMathOperator{\K}{\mathcal{K}}
\newcommand{\gamdmk}{\gamma_{mk}^{\dl}}
\newcommand{\gamdmkp}{\gamma_{mk'}^{\dl}}
\newcommand{\gamuml}{\gamma_{ml}^{\ul}}

\newcommand{\Wm}{\mathcal{W}_m^{\dl}}

\newcommand{\Ntx}{N}

\newcommand{\SIm}{\sigma^2_{\mathtt{SI},m}}
\newcommand{\betamkd}{\zeta_{f_{mk}}^{\dl}}

\newcommand{\betakldu}{\zeta_{h_{kl}}^{\mathtt{du}}}

\newcommand{\Pbhm}{P_{\mathtt{bh},m}}
\newcommand{\PUfix}{P_{\mathtt{U,fixed}}}

\newcommand{\SINRkdl}{\mathtt{SINR}_k^\dl}
\newcommand{\SINRelul}{\mathtt{SINR}_{l}^\ul}

\newcommand{\gmkd}{\qf_{mk}^{\dl}}

\newcommand{\hgmkd}{\hat{\qf}_{mk}^{\dl}}

\newcommand{\hgmlu}{\hat{\qg}_{ml}^{\ul}}

\newcommand{\gmlu}{\qg_{ml}^{\ul}}

\newcommand{\tgmkd}{\tilde{\qf}_{mk}^{\dl}}

\newcommand{\betamlu}{\zeta_{g_{ml}}^{\ul}}

\newcommand{\tgmlu}{\tilde{\qg}_{ml}^{\ul}}
\newcommand{\Ulul}{{\rm{U}}_l^\ul}
\newcommand{\Ukdl}{{\rm{U}}_k^\dl}

\newcommand{\betami}{{\zeta_{Q_{mi}}}}

\newcommand{\wikdl}{\qv_{ik}^{\dl}}
\newcommand{\wmkpdl}{\qv_{mk'}^{\dl}}

\newcommand{\wmlul}{\qv_{m\ell}^{\ul}}

\newcommand{\wmkdl}{\qv_{mk}^{\dl}}

\newcommand{\wmkdlmr}{\qv_{mk}^{\dl,\MR}}

\newcommand{\wmlulmr}{\qv_{m\ell}^{\ul,\MR}}

\begin{document}
\bstctlcite{IEEEexample:BSTcontrol}
 
\title{Cell-Free Full-Duplex Communication -- An Overview}

\author{Diluka Galappaththige, \IEEEmembership{Member, IEEE},  Mohammadali Mohammadi, \IEEEmembership{Senior Member, IEEE,}\\
Hien Quoc Ngo,~\IEEEmembership{Senior Member,~IEEE,} Michail Matthaiou,~\IEEEmembership{Fellow, IEEE,} and
Chintha Tellambura, \IEEEmembership{Fellow, IEEE}\\
\textit{(Invited Paper)}
\thanks{D. Galappaththige and C. Tellambura are with the Department of Electrical and Computer Engineering, University of Alberta, Edmonton, AB, T6G 1H9, Canada (e-mail: \{diluka.lg, ct4\}@ualberta.ca).} 
\thanks{M. Mohammadi, H. Q. Ngo, and M. Matthaiou are with the Centre for Wireless Innovation (CWI), Queen's University Belfast, BT3 9DT Belfast, U.K., email:\{m.mohammadi, hien.ngo, m.matthaiou\}@qub.ac.uk.}
 \vspace{-3mm}}

\maketitle  
\begin{abstract}Cell-free (CF) architecture and full-duplex (FD) communication are leading candidates for next-generation wireless networks. The CF framework removes cell boundaries in traditional cell-based systems, thereby mitigating inter-cell interference and improving coverage probability. In contrast, FD communication allows simultaneous transmission and reception on the same frequency-time resources, effectively doubling the spectral efficiency (SE). The integration of these technologies, known as CF FD communication, leverages the advantages of both approaches to enhance the spectral and energy efficiency in wireless networks. CF FD communication is particularly promising due to the low-power and cost-effective FD-enabled access points (APs), which are ideal for short-range transmissions between APs and users. Despite its potential, a comprehensive survey or tutorial on CF FD communication has been notably absent. This paper aims to address this gap in the literature. It begins with an overview of FD communication fundamentals, self-interference cancellation techniques, and CF technology principles, including their implications for current wireless networks. The discussion then moves to the integration and compatibility of CF and FD technologies, focusing on channel estimation, performance analysis, and resource allocation in CF FD massive multiple-input multiple-output (mMIMO) networks, supported by an extensive literature review and case studies. The potential of combining a sub-category of CF architecture—network-assisted CF technology—with FD technology is also explored, including a detailed case study on fundamentals, performance analysis, AP operation, and mode assignments. Finally, emerging CF FD paradigms, like millimeter wave (mmWave) communications, unmanned aerial vehicles, and reconfigurable intelligent surfaces, are discussed, highlighting existing contributions and unresolved issues.
\end{abstract}


\thispagestyle{empty}

\section*{Main Nomenclature}
\addcontentsline{toc}{section}{Nomenclature}
\begin{IEEEdescription}[\IEEEusemathlabelsep\IEEEsetlabelwidth{$V_1,V_2,V_3,V_4$}]
\fontsize{0.33cm}{0.4cm}\selectfont

\item[$5$G]     Fifth-generation
\item[$6$G]     Sixth-generation
\item[ADC]      Analog-to-digital converter
\item[AP]       Access point
\item[AWGN]     Additive white Gaussian noise
\item[BS]       Base station
\item[CF]       Cell-free
\item[CLI]      Cross-link interference 
\item[CPU]      Central processing unit
\item[CSI]      Channel state information
\item[DAC]      Digital-to-analog  converter
\item[DL]       Downlink 
\item[EE]       Energy efficiency 
\item[FD]       Full-duplex
\item[FDD]      Frequency-division duplex
\item[HD]       Half-duplex
\item[IBFD]     In-band full-duplex 
\item[IoT]      Internet of Things 
\item[IRS]      Intelligent reflecting surface 
\item[ISAC]     Integrated sensing and communication 
\item[I/Q]      In-phase/Quadrature phase 
\item[LoS]      Line-of-sight
\item[LSFD]     Large-scale fading decoding 
\item[ML]       Machine learning 
\item[MIMO]     Multiple-input multiple-output
\item[mMIMO]    Massive multiple-input multiple-output
\item[MMSE]     Minimum mean square error
\item[MRC]      Maximum ratio combining 
\item[MRT]      Maximum ratio transmission
\item[MSE]      Mean square error 
\item[NAFD]     Network-assisted full-duplex
\item[NLoS]     Non-line-of-sight
\item[NOMA]     Non-orthogonal multiple access
\item[OFDM]     Orthogonal frequency-division multiplexing
\item[PA]       Power amplifier
\item[PCA]      Principal component analysis
\item[PZF]      Partial zero-forcing
\item[QoS]      Quality-of-service
\item[RF]       Radio-frequency
\item[RI]       Residual interference
\item[RIS]      Reconfigurable intelligent surface 
\item[RV]       Random variable
\item[RZF]      Regularized zero-forcing
\item[SE]       Spectral efficiency 
\item[SI]       Self-interference
\item[SINR]     Signal-to-interference-plus-noise ratio
\item[SNR]      Signal-to-noise ratio 
\item[TDD]      Time-division duplex 
\item[UAV]      Unmanned aerial vehicle
\item[UE]       User equipment 
\item[UL]       Uplink 
\item[Wi-Fi]    Wireless fidelity
\item[ZF]       Zero-forcing

\end{IEEEdescription}

\section{Introduction} \label{Sec:intro}

\IEEEPARstart{T}{he}  fifth-generation (5G), beyond 5G, and sixth-generation (6G) wireless networks will be essential for all aspects of life, society, and industry, supporting technological applications like holographic telepresence, e-Health, smart environments, massive robotics, 3D unmanned mobility, augmented and virtual reality, and the Internet of Everything \cite{SamsungWhitePaper, Matthaiou:COMMag:2021}. These networks must meet diverse requirements in terms of spectral efficiency (SE), reliability, security, energy efficiency (EE), and latency \cite{Matthaiou:COMMag:2021}. Massive multiple-input multiple-output (mMIMO) systems, including co-located, distributed, and cell-free (CF) paradigms, have gained significant attention as key enabling technologies. In fact, the CF architecture, which combines the advantages of distributed antennas and co-located mMIMO, has attracted particular interest \cite{he2021cellfree}.

CF  has been developed to achieve high data rates, uniform quality-of-service (QoS), and ultra-high reliability by eliminating inter-cell interference through the absence of cell borders \cite{Zhang2019cellfree, he2021cellfree, Hien2017, Emil2020}. In conventional CF, numerous distributed access points (APs), connected to a central processing units (CPU) serve all users (UEs) cooperatively. However, conventional CF  systems face scalability issues, high fronthaul signaling, and computing complexity, making them problematic for large-scale networks \cite{Ngo:JPROC:2024,Demir2021book, Emil2020}. To address this, the user-centric approach has been deployed, where each user is supported by a small number of cooperative APs, reducing the load on the CPU. This topology enhances the EE and connectivity, ensuring consistent coverage and performance across the network \cite{Ngo:JPROC:2024,Demir2021book, Emil2020}.

Full-duplex (FD) operation, which can potentially double the SE of wireless networks by eliminating the need for separate uplink (UL) and downlink (DL) channels, has gained attention with advancements in self-interference (SI) cancellation techniques \cite{Riihonen2011, Sabharwal2012, Everett2014, Hong2014}. SI occurs when the transmitted signal from the FD node transmitter leaks to the receiver, often overwhelming the received signal of interest, which is transmitted from another node. Effective SI mitigation is thus critical to maximize the benefits of FD. Techniques such as propagation domain isolation, analog suppression, digital SI cancellation, and machine-learning (ML)/deep learning-based approaches have been developed for this purpose \cite{Choi2010, Everett2011, Radunovic2010, Cheng2012, Riihonen2009, Riihonen2011, Chun2010, Liu2015, Kim2015, Balatsoukas2018,  Guo2019, Kristensen2019, Sabharwal2012, Everett2014, Hong2014, Balatsoukas2020book, Tapio2021, Vaijayanti2022, Kolodziej2021RF, Kolodziej2021}. However, due to practical limitations, residual SI remains a critical factor in system performance. Despite these challenges, the FD technology is promising for next-generation wireless networks, potentially doubling the SE and addressing issues such as hidden terminals, congestion-related throughput loss, and large end-to-end delays \cite{Choi2010}.

\subsection{Existing Survey Papers and Organization}
Several separate surveys and tutorials exist for CF communication \cite{Demir2021book, Ammar2022, Giovanni2018, Zhang2020, Elhoushy2022, Zhang2019cellfree, Shuaifei2022, Kassam2023, Mohammadi:PROC} and FD communication \cite{Ashutosh2014, Zhang2016, Hong2014, Goyal2015,  Liu2015, Amjad2017, Sharma2018, Mohammadi2019, Alexandropoulos2022, smida2023full}. However, despite the brief discussion in \cite{Mohammadi2023} and an introduction to network-assisted CF-MIMO networks in~\cite{Mohammadi:TCOM:2024} no exclusive study focuses on the potential of consolidating CF and FD  technologies. This survey is the first to do so. Reference \cite{Mohammadi2023} highlighted the SE and EE enhancements  achieved by the integration of FD and CF, as well as corresponding challenges in future wireless networks. However, it did not fully explore recent advancements and integration challenges, leaving a significant gap in the literature. This study aims to fill that gap by comprehensively reviewing the amalgamation of FD and CF technologies.

The contributions of this paper are summarized as follows:
\begin{enumerate}
    \item 
The principles of FD communication are introduced, focusing on transceiver structures and the types of SI. Classical SI mitigation approaches, including passive suppression (propagation-domain SI cancellation) and active cancellation (analog- and digital-domain SI cancellations), are reviewed, along with recent ML-based SI mitigation methods. Additionally, the CF technology is briefly introduced, along with its implications for both existing and future wireless networks.

    \item The feasibility of merging CF and FD technologies is examined, focusing on channel estimation, performance analysis, and resource allocation in CF FD networks. An extensive literature review is provided, along with case studies using a general CF FD system configuration.

    \item The paper explores network-assisted FD (NAFD) CF networks, which use half-duplex (HD) transceivers to achieve virtual in-band FD (IBFD) communication. A comprehensive case study delves into the basics, performance analysis, and AP operation and mode assignments.

    \item
Finally, remaining challenges, open issues, and emerging trends are addressed, including a review of new CF FD paradigms like millimeter wave (mmWave) communications, unmanned aerial vehicle (UAV) communication, and reconfigurable intelligent surfaces (RISs), focusing on existing contributions and unresolved challenges.
\end{enumerate}

The remainder of this paper is as follows. Section~\ref{Sec:FD} presents an overview of FD technology, followed by a detailed discussion on the fundamentals of CF  in Section~\ref{Sec:cellfree}. In Section~\ref{Sec:Coexistence}, we explore the application of FD within CF  systems. Subsequently, Section~\ref{sec:NAFD} introduces NAFD CF, offering a unified approach to FD and HD CF. Finally, Section~\ref{sec:challenges} highlights the opportunities and challenges that CF FD pose for future 6G networks, with concluding remarks provided in Section~\ref{sec:conc}.

\textit{Notation:} We use bold upper case letters to denote matrices, and lower case letters to denote vectors. The superscripts $(\cdot)^H$ and $(\cdot)^{-1}$ stands for the conjugate-transpose (Hermitian) and matrix inverse, respectively;  $\mathbf{I}_N$ denotes the $N\times N$ identity matrix. The circular symmetric complex Gaussian distribution having variance $\sigma^2$ is denoted by $\mathcal{CN}(0,\sigma^2)$. Finally, $\mathbb{E}\{\cdot\}$ denotes the statistical expectation.   
\section{Full-Duplex Fundamentals} \label{Sec:FD}
Conventional wireless communication systems operate in HD mode, which uses orthogonal channels for end-to-end transmission \cite{Zhang2015}. In other words, a transceiver transmits and receives in non-overlapping time-slots, i.e., time-division duplex (TDD), or frequency-slots, i.e., frequency-division duplex (FDD), or in different orthogonal spectrum-spreading codes. This eliminates the possibility of strong SI from its transmission to reception \cite{Zhang2015}. Nevertheless, the unprecedented growth of wireless communication applications necessitates much higher SE for next-generation wireless technologies \cite{Matthaiou:COMMag:2021}. The FD operation, which allows for simultaneous transmission and reception over the same time and frequency resource blocks, is an appealing solution for meeting these requirements \cite{Ashutosh2014, Ngo:JSAC:2014, Hirley2020book, Zhang2016, Sharma2018,Mohammadi:TWC:2018}. However, the main challenge for FD systems is SI, which represents the leakage of an FD node's transmitted signal to the same node's receiver antennas. Suppose this SI can be canceled or suppressed to a sufficiently low level to allow proper detection of the low-power intended received signal from the other transceivers. In that case, FD communication could theoretically double the achievable SE \cite{Ashutosh2014, Hirley2020book}.

The SI is typically several orders of magnitude higher than the desired signal since the latter crosses a longer distance than the SI. For instance, consider two transceivers ($T_1$ and $T_2$) separated by \qty{500}{\m}; the desired signal from the $T_1$'s transmitter to the $T_2$'s receiver is attenuated by approximately \qty{120}{\dB}. If the isolation between the transmit and receive paths of the same transceiver is \qty{15}{\dB}, the SI would be \qty{105}{\dB} higher than the desired signal \cite{Ngoc2017book}. The massive difference in power levels between the SI and the desired signal further increases with the distance between the transceivers. This has historically limited the performance of FD communications. Thus, it has been widely believed that FD radios do not work. However, efficient SI cancellation techniques are now available, catalyzing research effort into FD technology as a wireless enabler beyond 5G \cite{Ashutosh2014, Hirley2020book, Zhang2016, Sharma2018}.

\begin{figure}[!t]\centering \vspace{0mm}
    \def\svgwidth{240pt} 
    \fontsize{8}{8}\selectfont 
    \graphicspath{{Figures/}}
    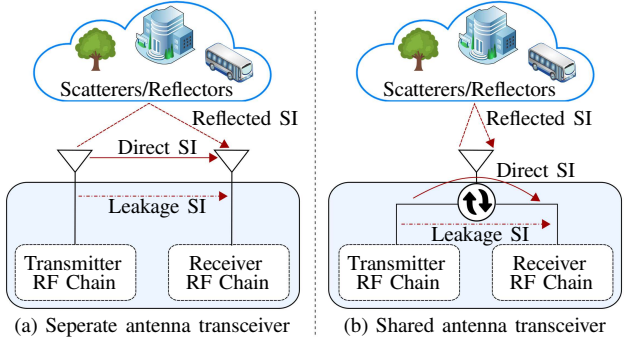 \vspace{0mm}
    \caption{The SI in separate antenna and shared antenna architectures.}\vspace{0mm} \label{fig:Tranceivers}
\end{figure}

There are two types of IBFD transceiver architectures: (i) separate antenna architectures, which use separate antennas for transmission and reception, and (ii) shared antenna architectures, which use the same antennas for simultaneous transmission and reception (Fig.~\ref{fig:Tranceivers}) \cite{Ngoc2017book}. In the latter case, a circulator separates the transmission and reception paths. Regardless of the transceiver architecture, i.e., separate antennas or shared antennas, there are three distinct types of SI (Fig.~\ref{fig:Tranceivers}) \cite{Debaillie2015}, i.e., (i) Leakage SI, (ii) Direct SI, and (iii) Reflected SI. The cumulative SI at the receiver, $x_{\rm{SI}}(t)$, can be mathematically expressed as
\begin{eqnarray} \label{eqn:SI_signal}
    x_{\rm{SI}}(t) &=& \underbrace{h_{\rm{SI}}^{\mathtt{l}} x_t(t-\tau_l)}_{x_{\rm{SI}}^{\mathtt{l}}(t) } 
     + \underbrace{h_{\rm{SI}}^{\mathtt{d}} x_t(t-\tau_d)}_{x_{\rm{SI}}^{\mathtt{d}}(t) } \nonumber \\
     &&+ \underbrace{\sum_{m=1}^M h_{{\rm{SI}},m}^{\mathtt{r}} x_t(t-\tau_m)}_{x_{\rm{SI}}^{\mathtt{r}}(t)}, 
\end{eqnarray}
where $x_{\rm{SI}}^{\mathtt{l}}(t)$, $x_{\rm{SI}}^{\mathtt{d}}(t)$, and $x_{\rm{SI}}^{\mathtt{r}}(t)$ represent the leakage SI, direct SI or spillover, and reflected SI, respectively, while $h_{\rm{SI}}^{\mathtt{l}}$, $h_{\rm{SI}}^{\mathtt{d}}$, $h_{{\rm{SI}},m}^{\mathtt{r}}$ denote the respective SI channels. Here, $\tau_l$, $\tau_d$, and $\tau_m$ represent the associated delays and $M$ is the number of multi-path reflection paths from the environment. In addition, $x_t(t)$ is the transmitted signal at the transmitter antenna. In particular, to generate $x_t(t)$, the digital-to-analog converter (DAC) in the transmitter first converts the baseband data/signal, $x_b[n]$, to its analog version, $x_b(t)$, i.e., the baseband signal. The transmitter then upconverts $x_b(t)$ to the RF signal, $x_u(t)$, and passes it to the power amplifier (PA) for amplification, resulting in the signal $x_t(t)$. 

\subsubsection{Leakage SI} This occurs on-chip or on-board in dense antenna integration and shared-antenna designs due to circulator leakage or cross-talk (e.g., due to imperfect antenna matching). This SI can be accurately characterized offline (i.e., in an anechoic chamber) and, thus, calibrated during the system design \cite{Ashutosh2014}.

\subsubsection{Direct SI} This comprises the signal propagating directly from the IBFD terminal’s transmit antennas to its own receive antennas, especially in separate antenna designs. This also occurs in shared antennas due to antenna impedance mismatches. These channels are usually modeled as line-of-sight (LoS) dominated channels, e.g., Ricean  \cite{Kim2015, Kim2021}. Due to the short distance of the direct link between the transmit and receive antennas, the direct SI power could exceed the desired signal power by up to \qty{104}{\dB}  in a wireless fidelity (Wi-Fi) system \cite{Ashutosh2014, Sahai2011 }. 

\subsubsection{Reflected SI} This is typically due to the non-LoS (NLoS) reflections of the transmitted signal from the external environment, such as nearby objects/scatters. Thus, the reflected SI depends on environmental effects that are changing and are inherently unpredictable. Such multi-path reflection leads to frequency-dependent/selective SI channels. These channels can be modeled empirically or analytically. The empirical models of SI channels derived from measurements are only efficient and accurate in environments with the same specific characteristics as the measurements \cite{Xiangyu2014, Chen2018, Yongyu2017}. The analytical channel models are more attractive than empirical models, and the geometry-based statistical channel model is one of the most commonly used analytical channel models \cite{Cezary2015, Liberti1996, Yifan2003, Wu2019}. These models are defined by the spatial location of the transmitter, receiver, and scatterers, which are described by the area form and spatial density of their occurrence. Typically, the reflected SI outpowers the desired received signal by \qtyrange{15}{100}{\dB} \cite{Margetts2012, Sahai2011}. The frequency selectivity and long propagation delay of reflected SI make the analog-circuit-domain cancellation much more challenging. 

The total SI due to these three sources of SI can be substantial. Hence,  effective SI cancellation techniques are indispensable for achieving IBFD communication.

\begin{figure}[!t]\centering \vspace{0mm}
    \def\svgwidth{200pt} 
    \fontsize{8}{8}\selectfont 
    \graphicspath{{Figures/}}    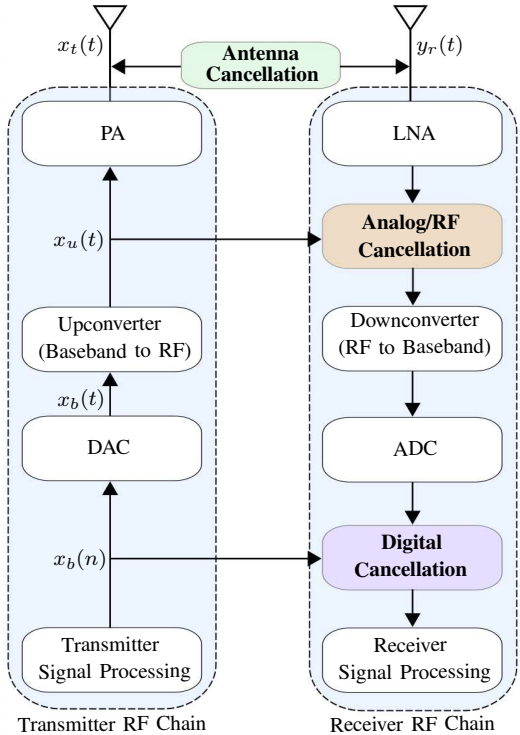 \vspace{0mm}
    \caption{A generic block diagram of an IBFD transceiver.}\vspace{0mm} \label{fig:FD_generic_block_diagram}
\end{figure}

The SI cancellation can occur at various points along the transmitter and receiver chains (Fig.~\ref{fig:FD_generic_block_diagram}). For example, replicating reference signals in the transmitter chain and subtracting the modified reference signal(s) in the receiver chain can remove multi-path SI. Here, additional signal delaying cancels the frequency-dependent SI, while attenuation and phase-shifting cancel the frequency-independent SI \cite{Debaillie2015}. 

Selecting a reference signal from the transmitted signal is critical because it must capture the information and distortions caused by RF imperfections, such as phase imbalance, gain, and transmitter-generated noise and distortion. For example, the analog-to-digital converter (ADC) dynamic range will be a significant bottleneck if this is done in the digital domain at the baseband, i.e., after the ADC. In particular, the strong SI is scaled along with the input to the ADC to match their respective dynamic ranges. From the classical rule of thumb for a \qty{10}{bit} ADC, the resultant quantization noise is $6.02 \times 10 + 1.76 = \qty{61.96}{\dB}$  lower than the signal at the input of the ADC \cite{Ngoc2017book}. If the SI is \qty{100}{\dB} higher than the desired signal, the quantization noise will be approximately \qty{38}{\dB} higher. Therefore, even if the SI is completely canceled at the ADC output, the receiver will not be able to process the desired signal.

On the other hand, the dominant nonlinearities in the PA processes cannot be eliminated if a reference signal is copied before it encounters the PA. Similarly, eliminating the SI early in the receiver's RF chain is advantageous because it reduces the demands on the receiver's front-end linearity, ability to handle large signals, and the resolutions required for the ADC. This highlights the significance of implementing SI cancellation near the antenna and, thus, choosing a reference signal as close to the antenna as possible is preferable. However, additional SI cancellation at the digital baseband is still required, especially to cancel the frequency-dependent SI. Furthermore, the transmitter impairments are significant compared to the received desired signal and must be reduced. Simply reducing the SI using the known transmitted symbols can result in a large residual SI. Hence, designing SI cancellation techniques is the most essential element of implementing the revolutionary FD paradigm \cite{Riihonen2011}.

\subsection{SI Mitigation}
Various SI cancellation/suppression schemes have been proposed for FD systems \cite{Choi2010, Everett2011, Radunovic2010, Cheng2012, Riihonen2009, Riihonen2011}. The proposed methods employ a combination of antenna (i.e., propagation-domain), radio-frequency (RF) (i.e., analog-domain), and baseband (i.e., digital-domain) techniques, which are classified into two categories.
\begin{enumerate}
    \item \textit{Passive suppression:}
    The propagation domain SI cancellation methods are passive and avail of physical methods to increase the propagation loss of the SI signal \cite{Choi2010, Chun2010, Liu2015, Kim2015}. Examples of these techniques are antenna separation, antenna cancellation, directional antennas, cross-polarization, and other techniques. Environmental reflections limit the effectiveness of cancellation. The maximum cancellation achieved in the reflective room, for example, is approximately \qty{27}{\dB} lower than in the anechoic chamber \cite{Everett2014}. Therefore,  additional cancellation (i.e., \qtyrange{30}{70}{\dB}) is required in the analog and digital domains to reduce the SI signal level to the noise floor \cite{Liu2015, Kim2015}.

    \item \textit{Active cancellation:}
    SI cancellations in the analog and digital domains are active methods. Analog cancellation can suppress the high-power SI caused by the ADC. It necessitates training sequence-based approaches or adaptive interference cancellation \cite{Choi2010, Sabharwal2010}. Since the residual SI after the analog active cancellation is still the rate-limiting bottleneck, additional digital cancellation is required. Digital cancellation requires delays and phase shifts between the transmitted and received signals to remove linear and nonlinear residual SI components. Thus, SI channel estimation is necessary \cite{Katti2013}.
\end{enumerate}
Combining them is an option because no single analog or digital method can provide sufficient SI canceling capacity. Nonetheless, nonlinear distortions, non-ideal frequency responses of circuits, phase noise, and other inevitable hardware imperfections in practical FD transceivers may limit the SI cancellation, leaving a significant residual SI \cite{Katti2013, Chun2010}.

\subsubsection{Propagation-Domain SI Cancellation}
These techniques aim to separate the transmitter and receiver chains electromagnetically. In particular, they suppress the SI before it manifests in the receive chain circuitry by physically isolating or separating an FD node's transmit and receive antennas \cite{Sabharwal2012}. The primary benefit of performing SI suppression in the propagation domain is that the receiver hardware does not need to accurately process signals with a wide dynamic range \cite{Ashutosh2014}. Separate antenna designs achieve this through a combination of path loss, cross-polarization, and antenna directionality \cite{Everett2014, Khojastepour2012, Everett2012phdthesis, Khojastepour2011}, whereas shared antenna architectures use the circulator, e.g., commercially available circulators can achieve transmit-receive isolation of about \qty{20}{\dB} \cite{Ngoc2017book}. 

Separate antenna systems can increase the path loss between the transmit and receive antennas by spacing them apart and/or by placing absorptive shielding between them \cite{Everett2014, Khojastepour2011, Everett2012phdthesis}.  Although this is a simple technique, its effectiveness is limited by the device's form factor, i.e., the smaller the device, the less room for such an implementation.  Cross-polarization provides an additional mechanism for isolating the IBFD transmit and receive antennas. For example, an IBFD terminal that transmits only horizontally polarized signals and receives only vertically polarized signals can avoid interference between them \cite{Everett2014, Khojastepour2012}.  Antenna directionality, on the other hand, aims to avoid interference via directional transmit and/or receive antennas by aligning null directions \cite{Everett2014, Everett2012phdthesis}. 

Even if factors like placement sensitivity and device integration hinder the performance of these techniques in practice, they can still be quite effective at suppressing the direct SI \cite{Everett2014}. For example, reference \cite{Everett2014} achieves a SI suppression level of \qty{74}{\dB} using commercially available hardware in an anechoic chamber setting, which is a much more accurate representation of an outdoor environment with generally fewer reflections. However, the same design performs poorly in highly reflective indoor environments, only providing \qty{46}{\dB} suppression. This is because it is sensitive to  reflected SI, the channel characteristics of which are unknown when the system is designed. It is worth noting that the same issues arise in shared-antenna systems, i.e.,  a circulator can accurately suppress the direct SI. Yet, it cannot distinguish between the reflected SI and desired signal.

On the other hand, dealing with reflected SI caused by nearby multi-path scattering, necessitates channel-aware SI suppression methods. Transmit beamforming is one such channel-aware propagation-domain SI suppression scheme. It directs the IBFD terminal's multi-antenna transmit array by custumizing the  complex weights per antenna to zero the radiation pattern at each IBFD receive antenna \cite{Snow2011, Everett2012phdthesis, Riihonen2011Mitigation}.  However, this requires accurate knowledge of the direct and reflected SI channel gains and delays, which can be acquired either explicitly through channel estimation or implicitly through weight adaptation \cite{Ashutosh2014}.

We point out that the aforementioned channel-aware or channel-unaware propagation-domain SI suppression techniques have a critical weakness, i.e., they may accidentally suppress the desired signal when adjusting the IBFD transmit and/or receive patterns to suppress the SI \cite{Ashutosh2014}.

\subsubsection{Analog-Domain SI Cancellation}
These methods aim to suppress the SI in the analog receive-chain circuitry before the ADC. In particular, the main idea is to replicate a reference signal(s) in the transmitter chain, process it in the analog domain, and subtract the modified reference signal(s) in the receiver chain to cancel the SI. This suppression can happen before or after the downconverter and low-noise amplifier. Other possibilities include replicating the reference signal in the digital domain, adjusting the essential gain/phase/delay, and then converting it to the analog domain for SI cancellation \cite{Sabharwal2010, Sabharwal2012, Duarte2014}. 
Like the propagation domain counterpart, analog domain cancellation techniques can be channel-aware or channel-unaware. 
Channel-unaware techniques cancel only the direct SI, whereas channel-aware techniques aim to cancel both the direct and reflected SI \cite{Everett2014, Everett2011, Sabharwal2010, Choi2010}. 

The direct and reflected SI channels can be modeled in a narrowband transmission scenario as complex gains and delays between each transmit and receiver antenna pair. In this case, a single antenna IBFD transceiver with separate transmit and receiver antennas needs to adjust only a single scalar complex cancellation gain and a single delay. A channel-unaware scheme can do this once the system is designed or calibrated. On the other hand, a channel-aware design constantly adjusts this gain and delay to compensate for changes in the reflected SI channel \cite{Ashutosh2014, Sabharwal2010, Choi2010}.

The same techniques as in narrowband transmission can mitigate the direct SI if the antenna gain and phase responses are frequency-flat. However, analog-domain SI cancellation is much more challenging because the reflected SI is generally frequency-selective due to multi-path propagation. In particular, replicating and processing a reference signal in the analog domain necessitates the adaptation of an analog filter for each transmit-receive antenna pair \cite{Ashutosh2014}. Another option is to digitally process the reference signal before converting it and using it in the analog domain. This makes adaptive digital filtering for reflected SI cancellation possible, which is typically much simpler to implement in wideband orthogonal frequency-division multiplexing (OFDM) systems \cite{Duarte2014}. However, non-idealities in the receiver's analog circuit limit the cancellation accuracy.

\subsubsection{Digital-Domain SI Cancellation}
These methods use sophisticated digital signal processing techniques on the received signal to cancel the SI after the ADC. Working in the digital domain has the advantage of making sophisticated processing relatively easy. For example, although beamforming can be implemented in the analog domain, it is common practice to implement it digitally due to lower circuit complexity and power consumption \cite{Riihonen2011Optimal, Chun2010}. However, the ADC's dynamic range severely limits the amount of possible SI reduction. Thus, for digital-domain methods to succeed, the propagation-domain and/or analog-domain SI cancellation techniques must achieve sufficient SI suppression prior to the ADC. In particular, as the last line of defense against SI, these methods aim to cancel the residual SI left over from the propagation- and analog-domain approaches.

Since the transmitted signal deviates from the generated reference signal due to hardware impairments, multi-path fading, and other imperfections, subtracting the estimated signal rather than the clean signal may improve the digital cancellation capabilities significantly.  In practice, digital cancellation has two main components: (i) estimating the SI channel and (ii) applying the channel estimation to the known transmitted signal to generate digital samples for subtracting the SI from the received signal \cite{Jain2011}. Estimating the SI signal components, including leakage through the analog cancellation circuit and delayed reflections of the SI signal from the environment, is necessary to eliminate the residual SI power after analog cancellation \cite{Katti2013}. There are two types of residual SI: linear and nonlinear. The linear SI accounts for the vast majority of SI power and can be approximated using least-square and minimum mean square error (MMSE)-based techniques \cite{scharf1991statistical}. On the other hand, nonlinear SI results from nonlinear distortions in imperfect analog cancellation circuits \cite{scharf1991statistical}. In particular, for high SI cancellation in the digital domain, the nonlinearity of the SI leakage channel must be precisely quantified.

\subsubsection{ML-based SI Mitigation}
Despite advances in classical SI cancellation techniques, perfect RF cancellation is challenging and expensive, resulting in residual SI signals at the receiver after the RF cancellation stages, i.e., propagation-domain and analog-domain cancellations \cite{Balatsoukas2020book}. Theoretically, this residual SI can be readily canceled in the digital domain since a known transmitted baseband signal causes it. Unfortunately, this is not practically feasible, as various transceiver nonlinearities, such as baseband nonlinearities (e.g., DAC and ADC), in-phase/quadrature-phase (I/Q) imbalance, phase noise, and PA nonlinearities, contaminate the SI signal \cite{Alexios2014, Korpi2014, Sahai2013, Syrjala2014, Korpi2017}. To fully cancel the SI to the level of the receiver noise floor, complicated nonlinear cancellation algorithms, often based on polynomial expansions, are necessary. For example, a parallel Hammerstein model is a commonly used nonlinear SI cancellation method that accounts for both PA nonlinearities and I/Q imbalance \cite{Korpi2017}.

Although polynomial models perform well in practice, their implementation complexity is often significant since the number of estimated parameters grows rapidly with the maximum considered nonlinearity order, whilst a large number of nonlinear basis functions must also be computed \cite{Korpi2017, Balatsoukas2020book}. On the other hand, principal component analysis (PCA), which can identify the most significant nonlinearity components of a parallel Hammerstein model, is an effective complexity-reduction technique.  However, PCA-based approaches require multiplying the transmitted digital baseband samples by a transformation matrix to obtain the SI cancellation signal,  thereby introducing additional complexity. Furthermore, the high-complexity PCA a procedure must be re-run whenever the SI channel undergoes significant change.

The most recent technologies that have received attention in academia and industry as a remedy for compensating for nonlinear effects in communications systems are ML and deep learning \cite{Raj2018, Balatsoukas2020book, Vaijayanti2022}. They have also been used for SI cancellation in FD radios \cite{Balatsoukas2020book, Vaijayanti2022, Balatsoukas2018, Tapio2021, Kolodziej2021RF, Kolodziej2021, Guo2019, Kristensen2019}. For example, due to their extensive nonlinear modeling capabilities, neural network based solutions for digital SI cancellation in FD radios have provided a favorable trade-off between computational complexity and SI cancellation performance \cite{Balatsoukas2020book}. The basic concept is to estimate the linear and nonlinear SI separately. In particular, the linear SI is first estimated using the least squares estimation method and then subtracted from the received signal. The remaining signal, which represents the nonlinear SI, is then estimated using ML or deep learning techniques and subtracted from the received signal for final SI cancellation \cite{Balatsoukas2020book, Vaijayanti2022}.

It is worth noting that ML-based techniques for SI mitigation operate via training models to more accurately predict and remove the SI signal than traditional methods \cite{Balatsoukas2020book}. These models can adapt to complex and dynamic environments, learning to model the nonlinearities and other factors that make SI cancellation challenging. Since these approaches operate in the digital domain (usually after signal ADC), they are classified as part of digital SI cancellation methods \cite{Balatsoukas2020book}.

For more information on the fundamentals of FD communication and SI cancellation, including classical and ML-based techniques, interested readers can refer to \cite{Mohammadi2023} and the references therein.

\section{Cell-Free Fundamentals}\label{Sec:cellfree}
The advent of 5G cellular technology has significantly improved data rates and traffic volumes compared to previous cellular technologies, thus reducing data transfer latency \cite{wong2017book}. However, it is essential to note that the benefits of these enhancements are more pronounced for users in proximity to cell centers. In contrast, users at the cell edges may still experience limitations due to inter-cell interference and handover issues inherent in cellular architectures. Therefore, while 5G presents a promising future for mobile communication, further advancements are needed to address these limitations and ensure that all users can benefit equally from this and future technologies.

In conventional cellular systems, the data source does not directly communicate with the user, which would require a high transmit power due to the user's distance. Instead, the data is sent to a nearby base station (BS) using low power, which then forwards it to a BS closer to the user. This approach allows for effective spatial reuse of the frequency spectrum and helps manage data traffic volumes. However, to handle higher traffic volumes, cellular networks often deploy more BSs in a given area, known as BS densification \cite{Zhang2020}. Unfortunately, this also increases the inter-cell interference and the frequency of handovers, resulting in traffic congestion at the cell edges. Unlike cell-center users, who experience lower interference levels and higher data rates, cell-edge users often receive moderate data rates even in 5G networks. This is reflected in the \qty{95}{\percent}-likely user data rates, which ensure that \qty{95}{\percent} of users receive the required minimum performance but remain unsatisfactory in 5G networks \cite{Zhang2020}. To overcome these limitations, innovative solutions are needed to improve the cell-edge performance and provide better user experiences.

One potential solution to address the limitations of conventional cellular systems is to adopt a CF architecture where each user is connected to several APs, effectively treating the entire network as a single cell \cite{Venkatesan2007, Shamai2001, Caire2010, Simeone2008, Marsch2011book, Diluka2019, Diluka2021, Diluka2020}. This approach eliminates the need for cell boundaries and significantly reduces inter-cell interference and handover issues. However, it also introduces new challenges, such as the requirement for massive fronthaul signaling for channel state information (CSI) and data sharing and high computational complexity. Several versions of this solution have been studied, including network MIMO \cite{Venkatesan2007, Shamai2001, Caire2010}, distributed MIMO \cite{Simeone2008}, and coordinated multi-point \cite{Marsch2011book}. To reduce fronthaul signaling and computational complexity, the network can be divided into clusters containing neighboring APs, with each cluster exchanging CSI and data only among themselves \cite{Huang2009, Zhang2009, Marsch2008}. While this approach can provide some performance gains, it only addresses the interference and handover issues within the cluster boundaries, leaving the cluster edges susceptible to these limitations \cite{Osseiran2011}.

A user-centric network could provide a potential solution to these challenges by allowing each AP to collaborate with different sets of APs when serving different users, putting the user in control of deciding which set of APs is best for its service instead of the network \cite{Hien2017, Nayebi2017, Zhang2019cellfree}. In addition, an abundance of service antennas relative
to the number of users is deployed. This approach is CF, and it has recently garnered significant research attention from the communication community \cite{Hien2017, Nayebi2017, Zhang2019cellfree}. This paradigm combines the best aspects of network MIMO and mMIMO developed over the last decade  \cite{Demir2021book, Emil2017}. It involves a user-centric approach where APs cooperate to serve users based on their specific channel conditions, allowing for more efficient use of available resources and reducing inter-cell interference and handover issues.

CF  is a novel approach that combines three previously known components: the mMIMO physical layer, ultra-dense networks with many more total number of AP antennas than users, and coordinated multipoint methods \cite{Demir2021book}. By consolidating these components, CF achieves user-centric, scalable operation for large-scale deployments. This approach offers several advantages over conventional cellular networks \cite{Demir2021book}:
\begin{itemize}
    \item It achieves smaller signal-to-noise ratio variations across the coverage area. 

    \item  It can help manage interference through joint processing at multiple APs, a capability that is very challenging in  dense cellular networks. 

    \item It can benefit from the increased SNR due to coherent transmission, allowing APs with weaker channels to participate in the transmission rather than just the AP with the best channel. 
\end{itemize}

Moreover, when a large number of co-located antennas are used at an AP, two important properties emerge in CF networks: (i) Channel hardening and (ii) Favorable propagation \cite{Demir2021book}. Channel hardening occurs when beamforming transforms a fading multiantenna channel into a nearly predictable scalar channel \cite{Demir2021book, Chen2018ChannelHardening, Zhang2019cellfree, Polegre2020}. In other words, small-scale fading effects disappear, and time/frequency-selective channels operate essentially as additive white Gaussian noise (AWGN) channels. This simplifies resource allocation as there is no need to adjust power allocation or scheduling to small-scale fading variations. Favorable propagation implies that the channel vectors of the users are almost orthogonal. This happens due to the law of large numbers and is beneficial because there will be little interference leakage between users. In principle, these properties result from the law of large numbers \cite{Demir2021book, Chen2018ChannelHardening, Zhang2019cellfree, Polegre2020}. In CF, multiple AP antennas contribute to these properties. However, due to the relatively smaller number of antennas per AP compared to cellular mMIMO, a lower degree of channel hardening is expected compared to cellular networks. On the other hand, CF networks are expected to provide favorable propagation between relatively distant users, but not among those who are close together.


Since the fundamentals of CF  have been well established, interested readers can refer to \cite{Demir2021book,Ngo:JPROC:2024} and the references therein for more information.

\section{Cell-free Full-Duplex Communications}\label{Sec:Coexistence}

CF FD communication is an emerging technology that combines the advantages of FD  and CF  systems, to improve the SE and EE of wireless networks \cite{Tung2019, Wang2020, Nguyen2020, Datta2022, Mohammadi2022, Anokye2021, Nguyen2020Heap}. It is a practical and promising technology for beyond 5G networks, thanks to the low-power and low-cost FD-enabled APs.

Despite the potential benefits, there are several challenges pertaining to the design of CF FD systems, especially in radio resource allocation. First, residual SI is a significant problem in FD communication and can limit the system's potential performance gains. Second, compared to conventional FD cellular networks, many APs and legacy users in CF  can lead to strong inter-AP interference and  co-channel interference  due to UL transmission to DL users. Third, the additional number of APs in CF FD increases the network power consumption, which needs to be managed carefully.
Despite its potential, there have only been a few attempts in the literature to characterize the performance of CF FD communication \cite{Nguyen2020Heap, Tung2019, Anokye2021, Datta2021,  Yu2023, Dey2022, Datta2022, Nguyen2020, Mohammadi2022,  Deng2022,    Wang2020,    Gao2023}.

\subsection{Channel Estimation}\label{sec_channel_est}

Channel estimation is critical in any wireless network configuration because it improves performance, reliability, and security \cite{Lu2022}. In general, accurate CSI is required for various communication tasks such as decoding, beamforming, equalization, and resource allocation. Similarly, accurate CSI is essential for signal detection, interference/SI mitigation, beamforming, and improving security and privacy in CF FD systems. Furthermore, channel estimates assist nodes, i.e., APs/users, in detecting changes in signal characteristics caused by interference or malicious attacks and in taking appropriate measures to protect communication \cite{Lu2022}.

To estimate the CSI, wireless networks typically use pilot-based channel estimation and blind or semi-blind techniques \cite{kim2015wireless, Ozdemir2007}. Pilot symbols are sent regularly or interleaved with data symbols to aid in channel estimation. Blind techniques, on the other hand, use the statistical properties of the received signal to estimate the CSI without the use of pilot symbols. However, due to the lack of pilot symbols, they may not achieve higher levels of accuracy (e.g., measured in terms of mean square error (MSE)) \cite{kim2015wireless, Ozdemir2007}.

Although pilots are required for accurate channel estimation in wireless systems, they come at a cost. Pilots can reduce the system's SE because they limit the time available for data transmission \cite{Hassibi2003, Emil2016}. Besides, transmitting more pilots to improve estimation accuracy can increase the system's energy consumption \cite{Hassibi2003, Emil2016}. As a result, it is essential to strike a balance between channel estimation accuracy and SE and EE to optimize the overall performance of the wireless system \cite{Hassibi2003, Emil2016}. Nonetheless, accurate channel estimation remains a critical component for improving the performance of wireless systems.

To this end, several channel estimation methodologies have been presented in the literature for CF FD systems \cite{Nguyen2020Heap, Tung2019, Anokye2021, Chowdhury:TCOM:2022, Datta2021, Datta2022,  Mohammadi2022, Hu:ACCESS:2021}. In particular, reference   \cite{Tung2019} presented an orthogonal pilot-based channel estimation for a generic CF FD system. It assumes that each AP has two multi-antenna arrays for transmission and reception. However, both antenna arrays can operate in transmit and receiver modes depending on the system requirement. To this end, the channels between the APs' transmit antennas and the DL users, as well as the APs' receiver antennas and the UL users, are simultaneously estimated using the MMSE methodologies used in CF  systems \cite{Demir2021book}. Similarly, references \cite{ Chowdhury:TCOM:2022, Mohammadi2022, Hu:ACCESS:2021} also utilized orthogonal pilot sequences to estimate the UL and DL channels using MMSE estimation. 

References \cite{Datta2022, Datta2021, Anokye2021, Nguyen2020Heap} used a slightly different approach. They divide the channel training period into two slots to estimate the UL and DL channels. Specifically, DL users first transmit their assigned orthogonal pilot sequences while the transmit antenna arrays at each AP operate in receiver mode. The APs thus estimate the DL channels using the received pilot signal through MMSE estimation. Similarly, in the next slot, the UL users transmit their pilot sequences, and the APs use the received signals at their receiver antenna arrays to estimate the UL channels. Furthermore, references \cite{Anokye2021, Nguyen2020Heap} employed a pilot sharing scheme to reduce channel training overload. In \cite{Nguyen2020Heap}, the authors proposed a novel heap-based pilot assignment algorithm that can mitigate the pilot contamination's effects while reducing the computational complexity.

\begin{table*}
	\centering
	\caption{Summary of CF FD channel estimation literature.}\label{tabel:channel_estimation}
	\small
\begin{tabular}{|c|c|c|c|c|c|c|>{\centering\arraybackslash}m{8.5cm}|}
	\hline
	\multirow{2}{*}{\textbf{Lit.}} & \multicolumn{3}{c|}{\textbf{AP}} & 
	\multicolumn{3}{c|}{\textbf{UE}} & \multirow{2}{*}{\textbf{Methodology}}\\

	\cline{2-7}
	\centering
	&\textbf{FD}  &\textbf{UL} &\textbf{DL}  &\textbf{FD} &\textbf{UL} &\textbf{DL}  & \\
    \hline  \hline

\centering
    \cite{Datta2022} & \checkmark   & - & - & - & \checkmark & \checkmark & \multirow{4}{*}{\begin{tabular}[c]{@{}l@{}}UL and DL channels are separately estimated at the APs using \\TDD protocol and orthogonal pilot-based MMSE estimation. \end{tabular}    }\\
	\cline{1-7}

\centering
    \cite{Anokye2021} & \checkmark   & - & - & - & \checkmark & \checkmark & \\
	\cline{1-7}
    
\centering
    \cite{Nguyen2020Heap} &  \checkmark & - & - & - & \checkmark & \checkmark & \\
	\cline{1-7}

\centering
    \cite{Datta2021} & \checkmark & - & - & - & \checkmark & \checkmark & \\
	\hline

\centering
    \cite{Mohammadi2022} & \checkmark   & - & - & - & \checkmark & \checkmark & \multirow{4}{*}{\begin{tabular}[c]{@{}l@{}}UL and DL channels are simultaneously estimated at the APs  \\using orthogonal pilot-based MMSE estimation. \end{tabular}    } \\
	\cline{1-7}
 
\centering
    \cite{Tung2019} & \checkmark & - & - & - & \checkmark & \checkmark &  \\
	\cline{1-7}

\centering
    \cite{Hu:ACCESS:2021} & - & \checkmark & \checkmark & - & \checkmark & \checkmark & \\
	\cline{1-7}

\centering
    \cite{Chowdhury:TCOM:2022} & - & \checkmark & \checkmark & - & \checkmark & \checkmark & \\
	\hline

\end{tabular}
\end{table*}

\textbf{Case study and discussion:}
A numerical/simulation example is presented to investigate one of the CF-FD channel estimation methodologies.

\begin{figure}[!t]\centering \vspace{1mm}
    \def\svgwidth{250pt} 
    \fontsize{8}{8}\selectfont 
    \graphicspath{{Figures/}}    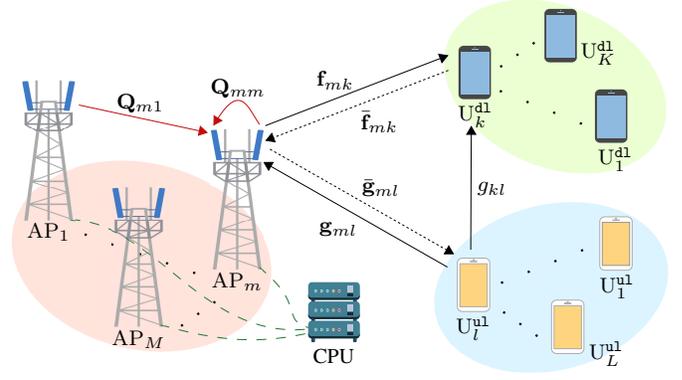 \vspace{0mm}
    \caption{A general CF FD system.}\vspace{-1em} \label{fig:SystemFigure}
\end{figure}

A CF FD system is considered consisting of $M$-FD APs, each with $N_t\geq 1$ transmit and $N_r\geq 1$ receive antennas, $K$-single antenna HD DL users, denoted by $\Ukdl$ for $k \in \mathcal{K}$, and $L$-single antenna HD UL users, denoted by $\Ulul$ for $l \in \mathcal{L}$ (Fig.~\ref{fig:SystemFigure}), where $\mathcal{K} \in \{1,\ldots, K\}$ and $\mathcal{L}\in \{1, \ldots, L\}$. The set of APs is  $\mathcal{M} \in \{1,\ldots,M \}$. A CPU is assumed to connect all APs via fronthaul/fronthaul links with sufficiently large capacities. All APs simultaneously serve all DL and UL users using the same time-frequency resources.  Each AP's transmit and receive antennas can operate in both transmit and receive modes. Each coherence interval includes two phases: (i) UL training and (ii) UL and DL data transmission. In particular, $\tau_p$ symbols from the coherence interval with $\tau_c$ symbols are used for UL channel estimation, while the remaining $(\tau_c-\tau_p)$ symbols are used for UL and DL data transmission.

Block flat-fading channel models are considered.  Thus, during each fading block, $\mathbf{f}_{mk} \in \mathbb{C}^{N_t \times 1}$ and  $\bar{\mathbf{f}}_{mk} \in \mathbb{C}^{N_r \times 1}$ denote the channel vectors from $\Ukdl$ to the $m$th AP transmit and receive antennas, respectively. The channels between $\Ulul$ and the the $m$th AP receive and transmit antennas are denoted by  $\mathbf{g}_{ml} \in \mathbb{C}^{N_r \times 1}$ and  $\bar{\mathbf{g}}_{ml} \in \mathbb{C}^{N_t \times 1}$, respectively. Moreover, $h_{kl} \in \mathbb{C}$ represents the channel between $\Ukdl$ and $\Ulul$, while $\mathbf{Q}_{mn} \in \mathbb{C}^{N_r \times N_t}$ denotes the channel between the $m$th AP transmit antennas and the $n$th AP receiver antennas.  All channels are assumed to be independent quasi-static Rayleigh faded, which remains constant during the coherence interval. A unified representation of all channels is given as
\begin{align}\label{eqn:chnl_model}
    \mathbf{a} = \zeta_{\mathbf{a}}^{1/2} \tilde{\mathbf{a}}, 
\end{align}
where $\mathbf{a} \in \{ \mathbf{f}_{mk}, \bar{\mathbf{f}}_{mk}, \mathbf{g}_{ml}, \bar{\mathbf{g}}_{ml}, {h}_{kl}\}$,   ${\zeta}_{\mathbf{a}}$ accounts for the large-scale path-loss and shadowing, and $\tilde{\mathbf{a}} \sim \mathcal{CN}\left(\mathbf{0},\mathbf{I} \right)$ captures the small-scale Rayleigh fading, which remains unchanged during one coherence interval.

During the channel training phase, each AP's transmit and receiver antennas operate in receiver mode. All UL and DL users simultaneously send their assigned pilot sequences to all APs. Let $\boldsymbol{\phi}_k^\dl \in \mathbb{C}^{\tau_p \times 1}$ and $\boldsymbol{\phi}_l^\ul \in \mathbb{C}^{\tau_p \times 1}$ be the pilot sequences transmitted by $\Ukdl$ and $\Ulul$, respectively. These pilot sequences are pairwisely orthogonal, i.e., $\left(\boldsymbol{\phi}_k^\dl \right)^{\rm{H}} \boldsymbol{\phi}_{k'}^\dl = 0$ for $k\neq k'$, $\left(\boldsymbol{\phi}_k^\dl \right)^{\rm{H}} \boldsymbol{\phi}_{l}^\ul = 0$, and $\left(\boldsymbol{\phi}_l^\ul \right)^{\rm{H}} \boldsymbol{\phi}_{l'}^\ul = 0$ for $l\neq l'$. This necessities $\tau_p \geq K+L$. Moreover, $\Vert \boldsymbol{\phi}_k^\dl \Vert^2 =1 $ for $k\in \mathcal{K}$ and $\Vert \boldsymbol{\phi}_l^\ul \Vert^2 =1 $ for $l\in \mathcal{L}$. Thus, the received signal at the $m$th AP transmit, $\mathbf{Y}_m^t \in \mathbb{C}^{N_t \times \tau_p}$, and receiver, $\mathbf{Y}_m^r \in \mathbb{C}^{N_r \times \tau_p}$, antennas are given as
\begin{subequations}
\begin{align}
    \mathbf{Y}_m^t \! &= \! \sqrt{\tau_p p_t} \sum_{i \in \mathcal{K}} \mathbf{f}_{mi} (\boldsymbol{\phi}_i^\dl )^{\rm{H}} + \sqrt{\tau_p p_t} \sum_{j \in \mathcal{L}} \bar{\mathbf{g}}_{mj} (\boldsymbol{\phi}_j^\dl )^{\rm{H}} + \mathbf{W}_m^t, \quad \label{eqn:tx_pilot} \\
     \mathbf{Y}_m^r \! &= \! \sqrt{\tau_p p_t} \sum_{i \in \mathcal{K}} \bar{\mathbf{f}}_{mi} (\boldsymbol{\phi}_i^\dl )^{\rm{H}} + \sqrt{\tau_p p_t} \sum_{j \in \mathcal{L}} \mathbf{g}_{mj} (\boldsymbol{\phi}_j^\dl )^{\rm{H}} + \mathbf{W}_m^r, \quad \label{eqn:rx_pilot}
\end{align}
\end{subequations}
where $p_t$ is the average pilot transmit power at each user, and $\mathbf{W}_m^t \in \mathbb{C}^{N_t \times \tau_p}$ and $\mathbf{W}_m^r \in \mathbb{C}^{N_r \times \tau_p}$ are the AWGN matrices having independent and identically distributed (i.i.d.) $\mathcal{CN} (0,\sigma_w^2)$ elements.

By projecting \eqref{eqn:tx_pilot} and \eqref{eqn:rx_pilot} onto $\boldsymbol{\phi}_k^\dl$ and $\boldsymbol{\phi}_l^\ul$, respectively, sufficient statistics for estimating $\mathbf{f}_{mk}$ and $\mathbf{g}_{ml}$ are given as 
\begin{subequations}
\begin{eqnarray}
     \mathbf{y}_{mk}^t  &=& \mathbf{Y}_m^t  \boldsymbol{\phi}_k^\dl = \sqrt{\tau_p p_t}  \mathbf{f}_{mk} + \mathbf{w}_{mk}^t, \quad \label{eqn:suf_tx_pilot} \\
     \mathbf{y}_{ml}^r &=& \mathbf{Y}_m^r \boldsymbol{\phi}_l^\ul = \sqrt{\tau_p p_t} \mathbf{g}_{ml} + \mathbf{w}_{ml}^r, \quad \label{eqn:suf_rx_pilot}
\end{eqnarray}
\end{subequations}
where $\mathbf{w}_{mk}^t = \mathbf{W}_m^t \boldsymbol{\phi}_k^\dl$ and $\mathbf{w}_{ml}^r = \mathbf{W}_m^r \boldsymbol{\phi}_l^\ul$. From \eqref{eqn:suf_tx_pilot},  the MMSE channel estimate of $\mathbf{f}_{mk}$ is obtained as
\begin{align}
    \hat{\mathbf{f}}_{mk} &= {\mathbb{E} \{ \mathbf{f}_{mk} (\mathbf{y}_{mk}^t) ^{\rm{H}}\} }\Big({ \mathbb{E} \{ \mathbf{y}_{mk}^t (\mathbf{y}_{mk}^t)^H  \}  }\Big)^{-1} \mathbf{y}_{mk}^t \nonumber\\
    &= c_{mk}^\dl \mathbf{y}_{mk}^t,
\end{align}
where 
\begin{eqnarray}
    c_{mk}^\dl = \frac{\sqrt{\tau_p p_t} \zeta_{f_{mk}} }{\tau_p p_t \zeta_{f_{mk}} + \sigma_w^2 }.
\end{eqnarray}
Similarly, from \eqref{eqn:suf_rx_pilot}, the MMSE channel estimate of $\mathbf{g}_{ml}$ is given as
\begin{align}
    \hat{\mathbf{g}}_{ml} &= {\mathbb{E} \{ \mathbf{g}_{ml}^{\rm{H}} \mathbf{y}_{ml}^r \} } \Big({ \mathbb{E} \{  \mathbf{y}_{ml}^r (\mathbf{y}_{ml}^r)^H \}  }\Big)^{-1} \mathbf{y}_{ml}^r \nonumber\\
    &= c_{ml}^\ul \mathbf{y}_{ml}^r,
\end{align}
where 
\begin{eqnarray}
    c_{ml}^\ul = \frac{\sqrt{\tau_p p_t} \zeta_{g_{ml}} }{\tau_p p_t \zeta_{g_{ml}} + \sigma_w^2 }.
\end{eqnarray}
Since $\mathbf{y}_{mk}^t$ and $\mathbf{y}_{ml}^r$ are Gaussian distributed,  $\hat{\mathbf{f}}_{mk} \sim \mathcal{CN} \left(\mathbf{0}, \gamdmk \mathbf{I}_{N_t} \right)$ and $\hat{\mathbf{g}}_{ml} \sim \mathcal{CN} \left(\mathbf{0}, \gamuml \mathbf{I}_{N_r} \right)$, where 
\begin{subequations}
\begin{eqnarray}
    \gamdmk &=& \frac{ \tau_p p_t \zeta_{f_{mk}}^2 }{\tau_p p_t \zeta_{f_{mk}} + \sigma_w^2 }, \\
    \gamuml &=& \frac{ \tau_p p_t \zeta_{g_{ml}}^2 }{\tau_p p_t \zeta_{g_{ml}} + \sigma_w^2 }.
\end{eqnarray}
\end{subequations}
Next, the channel estimation errors of $\mathbf{f}_{mk}$ and $\mathbf{g}_{ml}$ are respectively given as $\mathbf{e}_{mk}^\dl \triangleq \mathbf{f}_{mk} - \hat{\mathbf{f}}_{mk} \sim \mathcal{CN} \left(\mathbf{0}, (\zeta_{f_{mk}}-\gamdmk) \mathbf{I}_{N_t} \right)$ and $\mathbf{e}_{ml}^\ul \triangleq \mathbf{g}_{ml} - \hat{\mathbf{g}}_{ml} \sim \mathcal{CN} \left(\mathbf{0},(\zeta_{g_{ml}}-\gamuml) \mathbf{I}_{N_r} \right)$. Owing to the property of MMSE channel estimation, $\hat{\mathbf{f}}_{mk}$, $\mathbf{e}_{mk}^\dl$, $\hat{\mathbf{g}}_{ml}$, and $\mathbf{e}_{ml}^\ul$ are independent \cite{Demir2021book}.

The 3GPP UMi model is used to model the large-scale fading $\zeta_{\mathbf{a}}$ for $\mathbf{a} \in \{ {f}_{mk},  {g}_{ml}, {h}_{kl}, Q_{mn}\}$  with $f_c =  \qty{3}{\GHz}$ operating frequency \cite[Table B.1.2.1]{3GPP2010}. Moreover, the AWGN variance is modeled as $\sigma_w^2=10\log_{10}(N_0 B N_f)$ dBm, where $N_0=\qty{-174}{\dB m/\Hz}$, $B =  \qty{10}{\MHz}$ is the bandwidth, and $N_f = \qty{10}{\dB}$ is the noise figure. Moreover, the APs are uniformly distributed, while the users are randomly distributed over an area of $\num{400}\times\qty{400}{\m^2}$.

The quality of channel estimators is assessed by  the normalized MSE, which  is defined as 
\begin{eqnarray}
    \text{Normalized MSE} = \frac{\mathbb{E} \left\{  \left\Vert  \mathbf{a} - \hat{\mathbf{a}} \right \Vert^2 \right\}} {\mathbb{E} \left\{  \left\Vert  \mathbf{a} \right \Vert^2 \right\}}, 
\end{eqnarray}
where $\mathbf{a} \in \{\mathbf{f}_{mk}, \mathbf{g}_{ml}\}$ and $\hat{\mathbf{a}} \in \{ \hat{\mathbf{f}}_{mk}, \hat{\mathbf{g}}_{ml}\}$.


\begin{figure}[!t]\vspace{-1em}	
    \centering
    \includegraphics[width=0.46\textwidth]{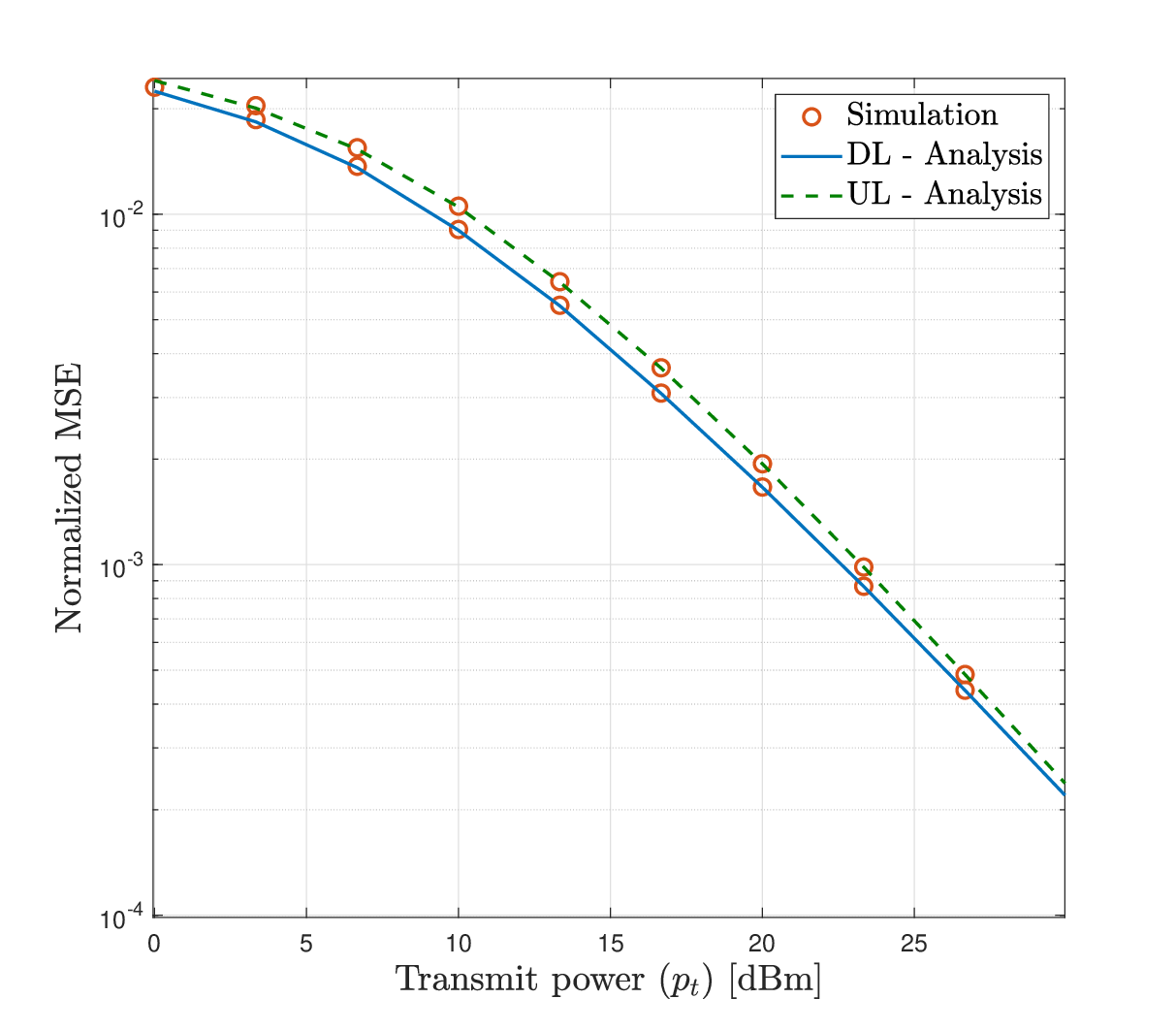}
    \vspace{-0.5em}
    \caption{Normalized MSE versus the per-user average  transmit power ($p_t$) for $M=16$, $N_t=N_r=4$, $K=L=2$, and $\tau_p=K+L$.}
    \label{MSE_ptx} \vspace{-0mm}
\end{figure}

Figure~\ref{MSE_ptx} plots the average normalized MSE of the DL and UL user channel estimates as functions of the average user pilot transmit power, $p_t$. As in conventional CF systems, the normalized MSE of both DL and UL user channels decreases with the per-user transmit power, improving the channel estimation quality. The key reason for this behavior is that using orthogonal pilot sequences prevents pilot contamination among DL users, DL-UL users, and vice versa. The transmit power thus improves the received signal quality at each AP. 


\begin{figure}[!t]\vspace{-1.5em}
    \centering
    \includegraphics[width=0.46\textwidth]{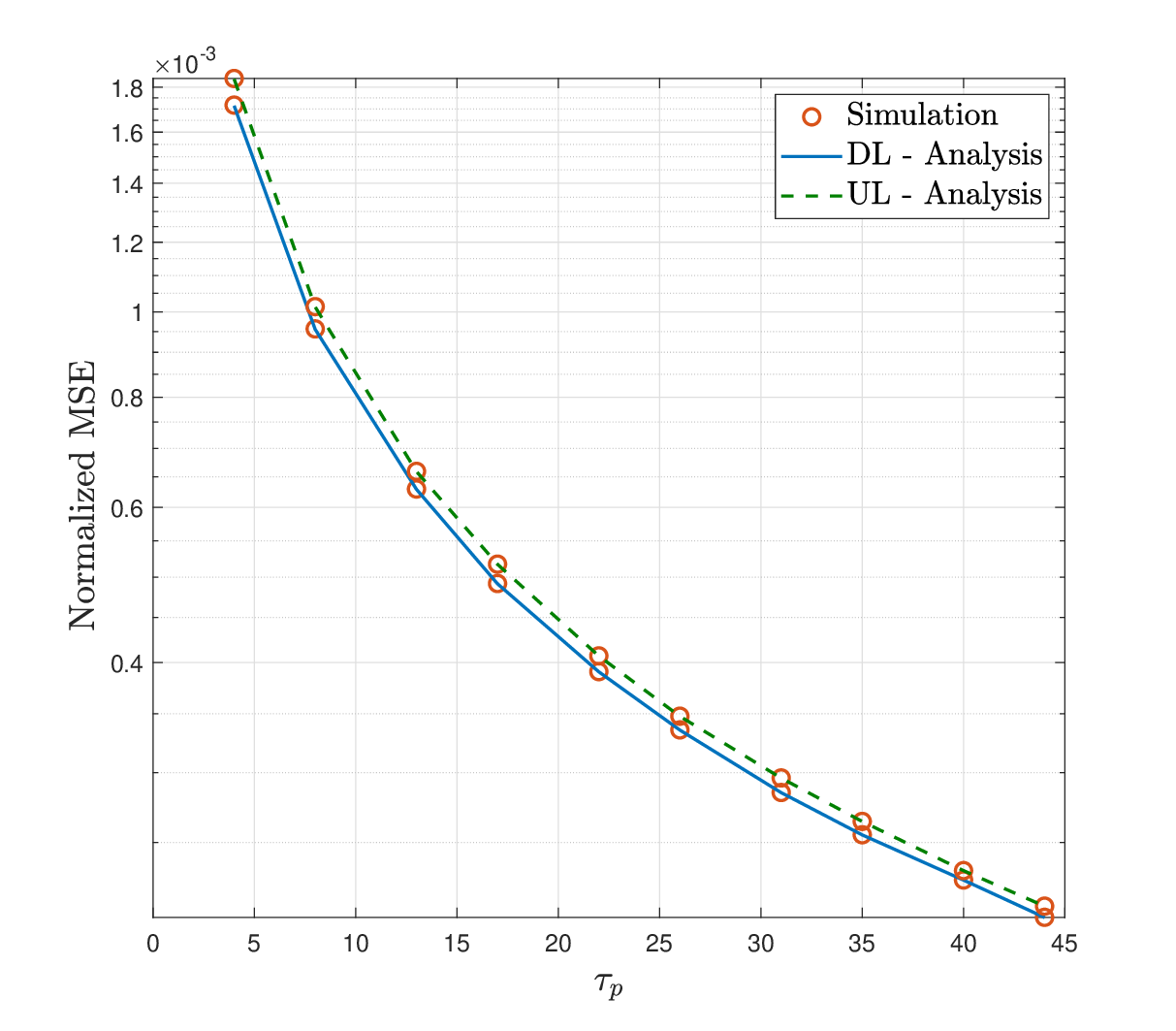}
    \vspace{-0.5em}
    \caption{Normalized MSE versus the pilot length ($\tau_p$) for $M=16$, $N_t=N_r=4$, $K=L=2$, and $p_t=\qty{20}{\dB m}$.}
    \label{MSE_tau} \vspace{-0mm}
\end{figure}

Figure~\ref{MSE_tau} shows the average normalized MSE of the DL and UL user channel estimates against the pilot length, $\tau_p$. The result reveals that longer pilot sequences result in more accurate DL and UL user channel estimates. For example, increasing the pilot sequence length from $\tau_p=\num{5}$ to $\tau_p=\num{30}$ can lower the normalized MSE by $\sim 10^{0.7}$ for the UL user channels.

The CF FD technical contributions can be categorized into two parts; (i) Performance analysis and (ii) Resource allocation.

\subsection{Performance Analysis}
Analytical performance evaluation is crucial in wireless networks, as it provides insights into the system potential and limitations. These approaches can determine throughput, delay, and error probabilities while considering practical transmission conditions and processing overheads. Developing such evaluations helps anticipate system requirements, computational costs, and accuracy for achieving practical goals, like the required QoS for applications. It also enables efficient resource utilization, helping to allocate resources efficiently. 

For instance, a few CF FD performance analysis frameworks are available \cite{Tung2019, Anokye2021}. Reference \cite{Tung2019} is the first paper to analyze a general CF FD system. The APs use conjugate beamforming and matched filtering to serve the DL and UL users, respectively, with the CSI acquired via UL training with orthogonal pilots transmitted from the users.  It derived the closed-form expressions for the UL and DL rates with a finite number of APs. Also, it proposed a simple power control method to mitigate the residual SI by investigating the effects of an infinite number of APs. Reference \cite{Anokye2021} investigated the SE/EE of a general CF FD system with low-resolution ADCs at the APs and DL users. The joint impact of residual SI, inter-AP interference, UL-to-DL interference, pilot contamination, and quantization noise was characterized by deriving closed-form solutions for the UL and DL rates. It was shown that increasing the number of receiving antenna arrays at the APs may compensate for the UL SE loss caused by quantization noise; however, increasing the transmit antennas is ineffective in compensating for the DL spectral loss. Furthermore, the effect of the ADCs' resolution on the operating region of the CF FD system and energy-spectral trade-offs are investigated.

\textbf{Case study and discussion:}
We will now examine the system performance for the configuration illustrated in Fig.~\ref{fig:SystemFigure}, utilizing the MMSE channel estimation outlined in Section \ref{sec_channel_est}.

\textit{DL Data Transmission:} The APs construct the precoders using the channel estimates from the training phase to transmit the information signals to the DL users. By applying conjugate beamforming \cite{Ngo2018}, the transmitted signal at the $m$th AP, $\mathbf{x}_m^\dl \in \mathbb{C}^{N_t \times 1}$, is given as
\begin{eqnarray}
    \mathbf{x}_m^\dl = \sqrt{p_d} \sum\nolimits _{k \in \mathcal{K}} \eta_{mk}^{1/2} \hat{\mathbf{f}}_{mk} q_k^\dl,
\end{eqnarray}
where $p_d$ is the maximum normalized transmit power at each AP, $\eta_{mk}$ for $m \in \mathcal{M}$ and $k \in \mathcal{K}$ is the transmit power allocation  coefficient at the $m$th AP for $\Ukdl$, and $q_k^\dl$ is the  symbols intended for $\Ukdl$, satisfying $\mathbb{E}\{\vert q_k^\dl\vert^2 \}=1$. Each AP must meet the average normalized power constraint, i.e., $\mathbb{E}\{\vert \mathbf{x}_m^\dl \vert^2 \} \leq p_d$, which entails the following per-AP power constraint:
\begin{eqnarray}
   N_t \sum\nolimits_{k \in \mathcal{K}} \eta_{mk} \gamdmk \leq 1, \quad \text{for} \quad m \in \mathcal{M}.
\end{eqnarray}
Next, the signal received at $\Ukdl$ is given as
\begin{eqnarray} \label{eqn_rx_DL_k}
    r_k^\dl &=& \sum\nolimits_{m\in \mathcal{M}} \mathbf{f}_{mk}^{\rm{H}} \mathbf{x}_m^\dl + \sum\nolimits_{l\in \mathcal{L}} h_{kl} x_l^\ul + w_k^\dl \nonumber \\
    &=& \underbrace{ \sqrt{p_d} \sum\nolimits_{m\in \mathcal{M}} \eta_{mk}^{1/2}  \mathbf{f}_{mk}^{\rm{H}} \hat{\mathbf{f}}_{mk}  q_k^\dl }_{\text{Desired signal }} \nonumber \\
   &&+  \underbrace{\sqrt{p_d} \sum\nolimits_{m\in \mathcal{M}} \sum\nolimits_{i\in \mathcal{K}_k} \eta_{mi}^{1/2}  \mathbf{f}_{mk}^{\rm{H}} \hat{\mathbf{f}}_{mi}  q_i^\dl }_{\text{Multi-user interference }}  \nonumber \\
   &&+ \underbrace{\sqrt{p_u} \sum\nolimits_{l\in \mathcal{L}} \varsigma_{l}^{1/2} h_{kl} q_{l}^\ul }_{\text{UL-to-DL interference }}  + \underbrace{w_k^\dl}_{\text{AWGN}},
\end{eqnarray}
where $\mathcal{K}_k \triangleq \mathcal{K}\backslash k$ and $w_k^\dl \sim \mathcal{CN}(0,\sigma_w^2)$ is the AWGN at $\Ukdl$. Moreover, $x_l^\ul$ is the transmitted signal at $\Ulul$ and given as $x_l^\ul = \sqrt{p_u\varsigma_{l}}  q_l^\ul$, where $p_u$ is the average UL user transmit power, $\varsigma_{l}$ is the power allocation coefficient at  $\Ulul$, and $q_l^\ul$ is the message symbol of $\Ulul$. 

\textit{UL Data Transmission:} UL and DL transmissions occur simultaneously at the same time-frequency block; $\Ulul$ transmits its signal to the APs, and the transmitted signal adheres to the average transmit power constraint, $\mathbb{E}\{\vert x_l^\ul \vert^2\} \leq p_u$, which results in the per-user power constraint as $0\leq \varsigma_{l} \leq 1$. The signal received  at the $m$th AP, $\mathbf{y}_m^\ul \in \mathbb{C}^{N_r \times 1}$, is given as
\begin{eqnarray}
    \mathbf{y}_m^\ul &=& \sum\nolimits_{l\in \mathcal{L}} \mathbf{g}_{ml} x_l^\ul + \sum\nolimits_{n\in \mathcal{M}} \mathbf{Q}_{mn} \mathbf{x}_n^\dl + \mathbf{w}_m^\ul \nonumber\\
    &=& \sqrt{p_u} \sum\nolimits_{l\in \mathcal{L}} \varsigma_{l}^{1/2}\mathbf{g}_{ml} q_l^\ul \nonumber\\
    &&\hspace{-1em}+ \sqrt{p_d} \sum\nolimits_{n\in \mathcal{M}}  \sum\nolimits_{k\in \mathcal{K}} \eta_{nk}^{1/2} \mathbf{Q}_{mn} \hat{\mathbf{f}}_{nk} q_k^\dl + \mathbf{w}_m^\ul, 
\end{eqnarray}
where $\mathbf{w}_m^\ul$ is the AWGN vector  with i.i.d. $\mathcal{CN} (0,\sigma_w^2)$ elements. To detect the message symbol transmitted by $\Ulul$, the $m$th AP multiplies the received signal $\mathbf{y}_m^\ul$ with the conjugate of the channel estimate, i.e., $\hat{\mathbf{g}}_{ml}^{\rm{H}}$, and then sends the resulting signal to the CPU.

Since the fronthaul links connect all APs to the CPU, the inter-AP interference can be viewed as the virtual residual interference (RI) of a virtual large BS with all transmit and receive antennas collocated \cite{Tung2019}. Hence,  residual inter-AP interference can be modeled similarly to the RI in FD systems \cite{Tung2019}. Thus,  Rayleigh fading is used to model the inter-AP interference channels, i.e., $\mathbf{Q}_{mn}$ is a matrix whose elements are i.i.d. $\mathcal{CN}(0, \theta_{\rm{SI}} \zeta_{Q_{mn}})$, where $\theta_{\rm{SI}}$ is the power of the RI at each AP after the SI suppression. 

The received signal from $\Ulul$ at the CPU is given as
\begin{eqnarray}\label{eqn_rx_UL_l}
    y_{ml}^\ul &=&  \sum\nolimits_{m \in \mathcal{M}} \hat{\mathbf{g}}_{ml}^{\rm{H}} \mathbf{y}_m^\ul \nonumber \\
    &=& \underbrace{ \sqrt{p_u} \sum\nolimits_{m\in \mathcal{M}} \varsigma_{l}^{1/2}  \hat{\mathbf{g}}_{ml}^{\rm{H}} \mathbf{g}_{ml}^{\rm{H}}  q_l^\ul }_{\text{Desired signal}} \nonumber \\
   &&+  \underbrace{\sqrt{p_u} \sum\nolimits_{m\in \mathcal{M}} \sum\nolimits_{j\in \mathcal{L}_l}\varsigma_{j}^{1/2} \hat{\mathbf{g}}_{ml}^{\rm{H}} \mathbf{g}_{ml}  q_j^\ul }_{\text{Multi-user interference}}  \nonumber \\
   &&+ \underbrace{\sqrt{p_d} \sum\nolimits_{m\in \mathcal{M}} \sum\nolimits_{n\in \mathcal{M}} \sum\nolimits_{k\in \mathcal{K}} \eta_{nk}^{1/2} \hat{\mathbf{g}}_{ml}^{\rm{H}} \mathbf{Q}_{mn} \hat{\mathbf{f}}_{nk} q_k^\dl }_{\text{Residual inter-AP interference }}  \nonumber\\
   &&+ \underbrace{ \sum\nolimits_{m\in \mathcal{M}}\hat{\mathbf{g}}_{ml}^{\rm{H}} \mathbf{w}_m^\ul}_{\text{Noise }},
\end{eqnarray}
where  $\mathcal{L}_l \triangleq \mathcal{L}\backslash l$.

\textit{DL SE Analysis:} The DL users are unaware of the instantaneous channel coefficients without DL pilot transmission, and thus, they rely on the statistical CSI for signal decoding \cite{Demir2021book}. To this end, the signal received  at $\Ukdl$ \eqref{eqn_rx_DL_k} is rearranged to decode by leveraging the statistical CSI as follows:
\begin{eqnarray}
    r_k^\dl &=&  \underbrace{ \sqrt{p_d} \mathbb{E} \left\{  \sum\nolimits_{m\in \mathcal{M}} \eta_{mk}^{1/2}  \mathbf{f}_{mk}^{\rm{H}} \hat{\mathbf{f}}_{mk} \right\}  q_k^\dl }_{\text{Desired signal $({\rm{DS}}_k^\dl)$}} \nonumber \\
    &&+\underbrace{ \sqrt{p_d} \!\left(  \sum\limits_{m\in \mathcal{M}} \!\! \eta_{mk}^{1/2}  \mathbf{f}_{mk}^{\rm{H}} \hat{\mathbf{f}}_{mk} \!-\! \mathbb{E}  \left\{ \! \sum\limits_{m\in \mathcal{M}} \!\! \eta_{mk}^{1/2}  \mathbf{f}_{mk}^{\rm{H}} \hat{\mathbf{f}}_{mk} \right\}  \right) q_k^\dl }_{\text{Beamforming gain uncertainty $({\rm{BU}}_k^\dl)$}} \nonumber \\
    &&+  \underbrace{\sqrt{p_d} \sum\nolimits_{m\in \mathcal{M}} \sum\nolimits_{i\in \mathcal{K}_k} \eta_{mi}^{1/2}  \mathbf{f}_{mk}^{\rm{H}} \hat{\mathbf{f}}_{mi}  q_i^\dl }_{\text{Multi-user interference $({\rm{MUI}}_k^\dl)$}}  \nonumber \\
    &&+ \underbrace{\sqrt{p_u} \sum\nolimits_{l\in \mathcal{L}} \varsigma_{l}^{1/2} h_{kl} q_{l}^\ul }_{\text{UL-to-DL interference $({\rm{UDI}}_k^\dl)$}}  + \underbrace{w_k^\dl}_{\text{AWGN}}.
\end{eqnarray}
The effective noise, i.e., the addition of ${\rm{BU}}_k^\dl$, ${\rm{MUI}}_k^\dl$, ${\rm{UDI}}_k^\dl$, and AWGN, can be treated as the worst-case independently distributed Gaussian noise in the sense of the central limit theorem \cite{Demir2021book}. This approximation gives a tight achievable rate bound  \cite{Demir2021book}. Assume that the users only use statistical CSI, hence, the  signal-to-interference-plus-noise ratio (SINR) at $\Ukdl$ is given as
\begin{align}\label{eqn_SINR_DL_k}
    \SINRkdl =\frac{ \vert {\rm{DS}}_k^\dl \vert^2 }{\mathbb{E} \{ \vert {\rm{BU}}_k^\dl \vert^2\} + \mathbb{E} \{ \vert {\rm{MUI}}_k^\dl \vert^2\} + \mathbb{E} \{ \vert {\rm{UDI}}_k^\dl \vert^2\} + \sigma_w^2 }. \quad
\end{align}
Next, by evaluating the expectation terms in \eqref{eqn_SINR_DL_k}, the closed-form solution of the SINR at $\Ukdl$ is given in \eqref{eqn_SINR_DL_k_close}:
\begin{align}\label{eqn_SINR_DL_k_close}
    &\SINRkdl = \nonumber\\
    &\frac{p_d N_t^2 \left(\sum_{m\in \mathcal{M}} \eta_{mk}^{1/2} \gamdmk \right)^2}{ p_d N_t \sum_{m \in \mathcal{M}} \sum_{i \in \mathcal{K}}\eta_{mi} \zeta_{f_{mk}} \rho_{mi}^\dl + p_u \sum_{l \in \mathcal{L}} \varsigma_{l} \zeta_{h_{kl}} +\sigma_w^2}.
\end{align}
The SE of $\Ukdl$ is thus given as
\begin{eqnarray}\label{eqn_DL_rate}
    \mathcal{S}_k^\dl = \frac{\tau_c-\tau_p}{\tau_c} \log_2 \left(1+ \SINRkdl \right),
\end{eqnarray}
where the pre-log factor  $(\tau_c-\tau_p)/\tau_c$ captures the effective portion of the coherence interval for the data transmission.

\textit{UL SE Analysis:}  The CPU also uses the statistical CSI to decode the data from the UL users. Rearranging the signal received from ${\rm{U}}_l^\dl$ in \eqref{eqn_rx_UL_l} and following the same methodology as in the DL SE analysis, the SE of ${\rm{U}}_l^\dl$ is expressed as
\begin{eqnarray}\label{eqn_UL_rate}
    \mathcal{S}_l^\ul = \frac{\tau_c-\tau_p}{\tau_c} \log_2 \left(1+ \SINRelul \right),
\end{eqnarray}
where $\SINRelul$ is ${\rm{U}}_l^\ul$ SINR at the CPU, whose closed-form solution is given in \eqref{eqn_SINR_UL_l_close} at the top of the next page.
\begin{figure*}
\begin{eqnarray}\label{eqn_SINR_UL_l_close}
    \SINRelul = \frac{p_u N_r^2 \left(\sum_{m\in \mathcal{M}} \varsigma_{l}^{1/2} \gamuml  \right)^2}{ p_u N_r \sum_{m \in \mathcal{M}} \sum_{j \in \mathcal{L}}\varsigma_{j} \zeta_{g_{mj}} \gamuml + p_d N_t \theta_{SI} \sum_{m \in \MM} \sum_{n \in \mathcal{M}} \sum_{k \in \mathcal{K}} \eta_{nk}  \zeta_{Q_{mn}} \gamuml \rho_{nk}^{d} +\sigma_w^2 \sum_{m\in \mathcal{M}} \gamuml }
\end{eqnarray}
\hrulefill
\end{figure*}

\subsection{Resource Allocation}
Spectrum, transmit power, bandwidth, and other resources are limited. Hence, sharing these resources among multiple users leads to resource allocation problems. Hence, optimal resource allocation is crucial from both theoretical and practical perspectives. Since wireless users may have different requirements and priorities, and wireless networks can be of various types (e.g., commercial networks such as Wi-Fi and LTE, sensor networks, and energy harvesting networks), there is no one-size-fits-all solution. Therefore, the type of resource allocation is application-dependent.  However, while some wireless application problems are amenable to standard resource allocation strategies, others need specialized solutions tailored to specific needs.

CF-FD resource allocation has its challenges,  particularly in managing SI and inter-AP interference, which cannot be entirely eliminated.   Consequently, the residual SI must be considered when assigning wireless resources (time, power, and frequency channels) to users. Resource allocation schemes have been developed to address this issue \cite{Yu2023, Dey2022, Datta2022, Datta2021, Nguyen2020, Mohammadi2022, Gao2023}.

Reference \cite{Nguyen2020} investigated a CF FD system for improved SE and EE. The authors proposed a SE and EE maximization problem by jointly optimizing a power control, AP-user association, and AP selection. Furthermore, the realistic power consumption model, which accounts for data transmission, baseband processing, and circuit operation, was used to characterize the EE performance. References \cite{Datta2021, Datta2022} studied the EE maximization of a CF FD system with limited-capacity fronthaul links via the DL and UL power control. In \cite{Datta2022},  a two-layered approach was employed, where it first formulated the optimization as a generalized convex program and then solved it either centrally or decentrally using the alternating direction method of multipliers. Reference \cite{Mohammadi2022} proposed a virtual FD approach for CF, where IBFD is virtually implemented by using the existing HD APs without SI-suppression hardware. It developed a mixed-integer problem to maximize the sum SE by jointly designing the AP mode assignment and power control.

Reference \cite{Yu2023} proposed a joint beamforming design for access and fronthaul links in a user-centric network with FD fronthaul. By accounting for the power consumed by SI cancellation at FD APs, the proposed optimization scheme maximized the network EE while ensuring fronthaul rate requirements. 
In \cite{Dey2022}, the UL/DL SE of an CF FD system was investigated with estimated effective DL channel, ADC/DAC impairments at the APs and the users, and spatially-correlated Ricean channels. The authors maximized the non-convex global EE metric using the block minorization-maximization technique, which decomposes the problem into several convex surrogate sub-problems. Reference \cite{Gao2023} investigated a FD CF non-orthogonal multiple access (NOMA)-assisted space-ground integrated network, which employs spectrum sharing between the satellite and terrestrial networks to improve the SE. The authors proposed a sum-rate maximization scheme that optimizes the power allocation factors of the NOMA DL, the UL transmit power, and both the satellite and AP beamformer. Reference \cite{Anokye:TWC:2023} studied the UL/DL transmit power minimization to maximize the SE and EE of a CF FD system over Ricean fading channels under SI and inter-AP interference at APs, UL-to-DL interference at DL users, and low-resolution ADCs at both APs and DL users. Table~\ref{tabel:CF_literature} provides a summary of the existing literature on CF FD.

\begin{table*}
	\centering
	\caption{Summary of CF FD performance analysis and resource allocation literature. }\label{tabel:CF_literature}
	\small
\begin{tabular}
{|C{1.8cm}|C{0.8cm}|c|c|c|C{1.7cm}|C{8.1cm}|}
	\hline
	\multirow{2}{*}{\textbf{Category}} & \multirow{2}{*}{\textbf{Lit.}} & \multicolumn{3}{c|}{\textbf{Setup}} & \multirow{2}{*}{\begin{tabular}[c]{@{}c@{}}\textbf{Channel} \\\textbf{Estimation}\end{tabular}}
	 & \multirow{2}{*}{\textbf{Technical Contribution}}\\

	\cline{3-5}
	\centering
	& &\textbf{AP}  &\textbf{UL UE} &\textbf{DL UE}  &    & \\
    \hline  \hline

\multirow{8}{*}{\begin{tabular}[c]{@{}c@{}}Performance \\analysis\end{tabular}}
  & \cite{Anokye2021}     &  FD   & HD & HD & Separate &  Closed-form solutions for UL/DL SE to characterize the joint effect of residual SI, inter-AP interference, UL-to-DL interference, pilot contamination, and quantization noise.   \\
	\cline{2-7}
    
\centering
    & \cite{Tung2019}     &   FD   & HD & HD & Simultaneous &  Closed-form expressions of the UL and DL achievable rates using conjugate beamforming with a finite number of APs. A power control method to mitigate the residual SI with a large number of APs.\\
	\cline{2-7}
 
\centering
     & \cite{Nguyen2020}    & FD    & HD & HD  & Heap-based pilot assignment & Maximizing the SE and EE by jointly optimizing power control, AP-user association, and AP selection under a realistic power consumption model. \\
	\hline

\centering
\multirow{20}{*}{\begin{tabular}[c]{@{}c@{}}Resource \\allocation\end{tabular}}
 & \cite{Datta2022}    & FD   & HD  & HD  & TDD-based orthogonal pilots  &  Weighted sum EE maximization via  DL and UL power control utilizing a two-layered approach with limited-capacity fronthaul links. \\
	\cline{2-7}

\centering
    & \cite{Mohammadi2022}    &  FD  & HD  & HD & Simultaneous orthogonal pilots & DL and UL SE maximization via a mixed-integer optimization problem that concurrently designs the AP mode assignment and power control with a virtual FD mode at APs. \\
	\cline{2-7}

\centering
    & \cite{Datta2021}    & FD   & HD & HD & TDD-based orthogonal pilots & Weighted sum EE maximization through DL and UL power control and a successive convex approximation framework with limited-capacity fronthaul links.  \\
	\cline{2-7}

\centering
    & \cite{Yu2023}    &  FD  & - & HD & - &  EE maximization of a user-centric network with FD fronthaul by designing beamforming for access and fronthaul links while ensuring fronthaul rate requirements. \\
	\cline{2-7}

\centering
    & \cite{Dey2022}    &  FD  & FD & & TDD-based orthogonal pilots &  Global EE maximization using the minorization-maximization technique, accounting for RF impairments at APs and UEs, ADC/DAC resolutions, and spatially-correlated Ricean channels.   \\
	\cline{2-7}

\centering
    & \cite{Gao2023}    & FD   & HD & HD & TDD-based orthogonal pilots   &  Sum-rate maximization of a CF FD NOMA-assisted space-ground integrated network by optimizing the NOMA DL power allocation factors, UL transmit power, and beamforming at both the satellite and APs. \\
	\cline{2-7}

\centering
    & \cite{Anokye:TWC:2023}    & FD   & HD & HD & TDD-based orthogonal pilots   &  UL/DL transmit power minimization over Ricean fading channels under SI and inter-AP interference at the APs, UL-to-DL interference at DL users, and low-resolution ADCs at both APs and DL users. \\
	\hline

\end{tabular}
\end{table*}


\textbf{Case study and discussion:} Power allocation is now presented for the CF FD system in Fig.~\ref{fig:SystemFigure}.  

\subsubsection{Transmit Power Control}
The spatial distribution of the (UL/DL) users causes the near-far effect, which affects the achievable rates. Nonetheless, max-min transmit power control, optimal for user fairness in mitigating the near-far effect, can provide uniform QoS to all users, as in conventional CF systems. To this end,  for a given realization of large-scale fading, the UL and DL power control coefficients can be found, i.e., $\eta_{mk}$ and $\varsigma_{l}$, that maximize the minimum of all UL and DL  user rates, respectively, under the relevant constraints. Since the  UL and DL transmissions occur concurrently, the multi-objective optimization technique must be invoked \cite{Bjornson2014}. Besides, since the rates in \eqref{eqn_DL_rate} and \eqref{eqn_UL_rate} are monotonically increasing functions of their arguments, i.e., $\SINRkdl$ and $\SINRelul$, they can be equivalently replaced with SINR terms in \eqref{eqn_SINR_DL_k_close} and \eqref{eqn_SINR_UL_l_close}, respectively. By introducing a common SINR, $\lambda$, for the UL and DL users and defining the slack variables $\mu_{mk} \triangleq \eta_{mk}^{1/2}$ and $\alpha_l \triangleq \varsigma_{l}^{1/2}$, a multi-objective optimization problem (MOOP) is formulated as follows:
\begin{subequations}\label{P1_prob}
    \begin{eqnarray}
        \mathbf{P1}:
        \underset {\beta_{mk}, \alpha_l}{\rm{max}} && (\lambda_d)^{w_d} (\lambda_u)^{w_u} =\lambda_c \label{P1_obj} \\
        \text{s.t}\hspace{0.7em} && \left(\SINRkdl \right)^{w_d} \left(\SINRelul \right)^{w_u} \geq \lambda_c,   \label{P1_sinr}\\
        && N_t \sum_{k \in \mathcal{K}} \beta_{mk}^2 \gamdmk \leq 1,  \label{P1_AP_power}\\
        &&  0 \leq \beta_{mk} \leq 1,   \label{P1_eta} \\
        && 0 \leq \alpha_l \leq 1,  \label{P1_beta}
    \end{eqnarray}
\end{subequations}
where $w_d$ and $w_u$ are the priorities assigned for the DL and UL users, respectively. The main SINR constraint \eqref{P1_sinr} in $\mathbf{P1}$ is quasi-concave \cite{Hien2017}. The under-laying optimization problem is also quasi-concave \cite{Hien2017}. Hence, $\mathbf{P1}$ is amenable to a bisection search, solving a sequence of convex feasibility problems in each iteration. \textbf{Algorithm \ref{Algo1}} gives the details. Moreover, $\mathbf{v}_k^\dl\triangleq \left[\sqrt{N_t}\mathbf{v}_{k1}^\dl, \sqrt{\frac{p_u}{p_d}}\mathbf{v}_{k2}^\dl, \frac{\sigma_w}{\sqrt{p_d}} \right] $ and $\mathbf{v}_l^\ul\triangleq \left[\sqrt{N_r}\mathbf{v}_{l1}^\ul, \sqrt{\frac{p_d N_t \theta_{\rm{SI}}}{p_u}}\mathbf{v}_{l2}^\ul, \frac{\sigma_w}{\sqrt{p_u}}\mathbf{v}_{l3}^\ul \right] $, where 
\begin{subequations}
    \begin{eqnarray}
        \mathbf{v}_{k1}^\dl &\triangleq& \left[\mu_{11} \sqrt{\zeta_{f_{1k}}\rho_{11}^\dl }, \ldots, \mu_{MK} \sqrt{\zeta_{f_{Mk}}\gamdmk } \right] \in \mathbb{R}^{MK },\qquad \\
        \mathbf{v}_{k2}^\dl &\triangleq& \left[\alpha_{1} \sqrt{\zeta_{h_{kl}}}, \ldots, \alpha_{L} \sqrt{\zeta_{h_{kL}}}\right] \in \mathbb{R}^{L },\qquad
    \end{eqnarray}
\end{subequations}
and
\begin{subequations}
    \begin{eqnarray}
        \mathbf{v}_{l1}^\ul &\triangleq& \left[\alpha_{1} \sqrt{\zeta_{g_{11}}\rho_{1l}^\ul }, \ldots, \alpha_{L} \sqrt{\zeta_{g_{ML}}\gamuml } \right] \in \mathbb{R}^{ML},\qquad \\
        \mathbf{v}_{l2}^\ul &\triangleq& \left[\mu_{11} \sqrt{\zeta_{Q_{11}} \rho_{1l}^\ul \rho_{11}^\dl },\ldots, \mu_{nk} \sqrt{\zeta_{Q_{mn}} \gamuml \rho_{nk}^\dl }, \right. \nonumber\\
         && \left. \qquad \ldots, \mu_{MK} \sqrt{\zeta_{Q_{MM}} \gamuml \gamdmk } \right] \in \mathbb{R}^{M^2K }, \\
        \mathbf{v}_{l3}^\ul &\triangleq& \left[\sqrt{\gamuml}, \ldots, \sqrt{\gamuml}\right] \in \mathbb{R}^{M }.\qquad
    \end{eqnarray}
\end{subequations}

\begin{algorithm}[t]
\caption{: Bisection algorithm for solving $\mathbf{P1}$.}\label{Algo1}
\begin{algorithmic}
    \renewcommand{\algorithmicrequire}{\textbf{Initialization:}}
    \renewcommand{\algorithmicensure}{\textbf{Repeat}}
    \Require Select the initial values of $\lambda_{\rm{min}}$ and $\lambda_{\rm{max}}$, where $\lambda_{\rm{min}}$ and $\lambda_{\rm{max}}$ define a range of $\mathbf{P1}$. Choose a tolerance $\epsilon > 0$. 
    \Ensure
    \State \textbf{Step 1}: Set $\lambda_c = (\lambda_{\rm{min}} + \lambda_{\rm{max}})/2$. 
    \State \textbf{Step 2}: Solve the  following convex feasibility problem:
    \begin{eqnarray}\label{eqn_feasible_pron}
    \begin{cases}
         & \Vert \mathbf{v}_k^\dl \Vert \leq \frac{N_t}{\sqrt{\lambda_c^{w_d}}} \left( \sum\limits_{m \in \mathcal{M}} \mu_{mk} \gamdmk \right), \quad \text{for} \quad k\in \mathcal{K}, \quad\\
          & \Vert \mathbf{v}_l^\ul \Vert \leq \frac{N_r}{\sqrt{\lambda_c^{w_u}}} \left( \sum\limits_{m \in \mathcal{M}} \alpha_{l} \gamuml \right), \quad \text{for} \quad l\in \mathcal{L}, \\
        & N_t \sum\limits_{i\in \mathcal{K}} \mu_{mi}^2 \rho_{mi}^\dl \leq 1,  \quad \text{for} \quad m\in \mathcal{M}, \\
        & 0 \leq \beta_{mk} \leq 1, \quad \text{for} \quad  m\in \mathcal{M} \quad \text{and}\quad k\in \mathcal{K}, \\
        &  0 \leq \alpha_{l} \leq 1,   \quad \text{for} \quad  l\in \mathcal{L}.
    \end{cases}
    \end{eqnarray}
    
    \State \textbf{Step 3}: If problem \eqref{eqn_feasible_pron} is feasible, then set $\lambda_{\rm{min}}= \lambda_c$, else set $\lambda_{\rm{max}}= \lambda_c$.
\end{algorithmic}
\textbf{Until} $\lambda_{\rm{max}} - \lambda_{\rm{min}} < \epsilon$. \\
\textbf{Output:} The optimal power allocation coefficients, $\eta_{mk}= \mu_{mk}^2$ for $m \in \mathcal{M}$ and $k \in \mathcal{K}$ and $\varsigma_{l}=\alpha_l^2$ for $l \in \mathcal{L}$. 
\end{algorithm}

Equal DL and UL priorities are set for max-min rates, i.e., $w_d=w_u=1$. Equal power allocation is set at the APs, i.e., $\eta_{mk}=1/K$ for $m\in \mathcal{M}$ and $k\in \mathcal{K}$. Moreover, full power transmission at the UL users, i.e., $\varsigma_{l}=1$ for $l\in \mathcal{L}$, is considered to evaluate the performance of our proposed power control method (\textbf{Algorithm~\ref{Algo1}}). Monte-Carlo simulated user SEs are also superimposed on each curve to validate the derived analytical expressions. Moreover, let  $\theta_{SI}=\qty{-70}{\dB}$.

\begin{figure}[!t]\vspace{-2em}	
    \centering
    \includegraphics[width=0.48\textwidth]{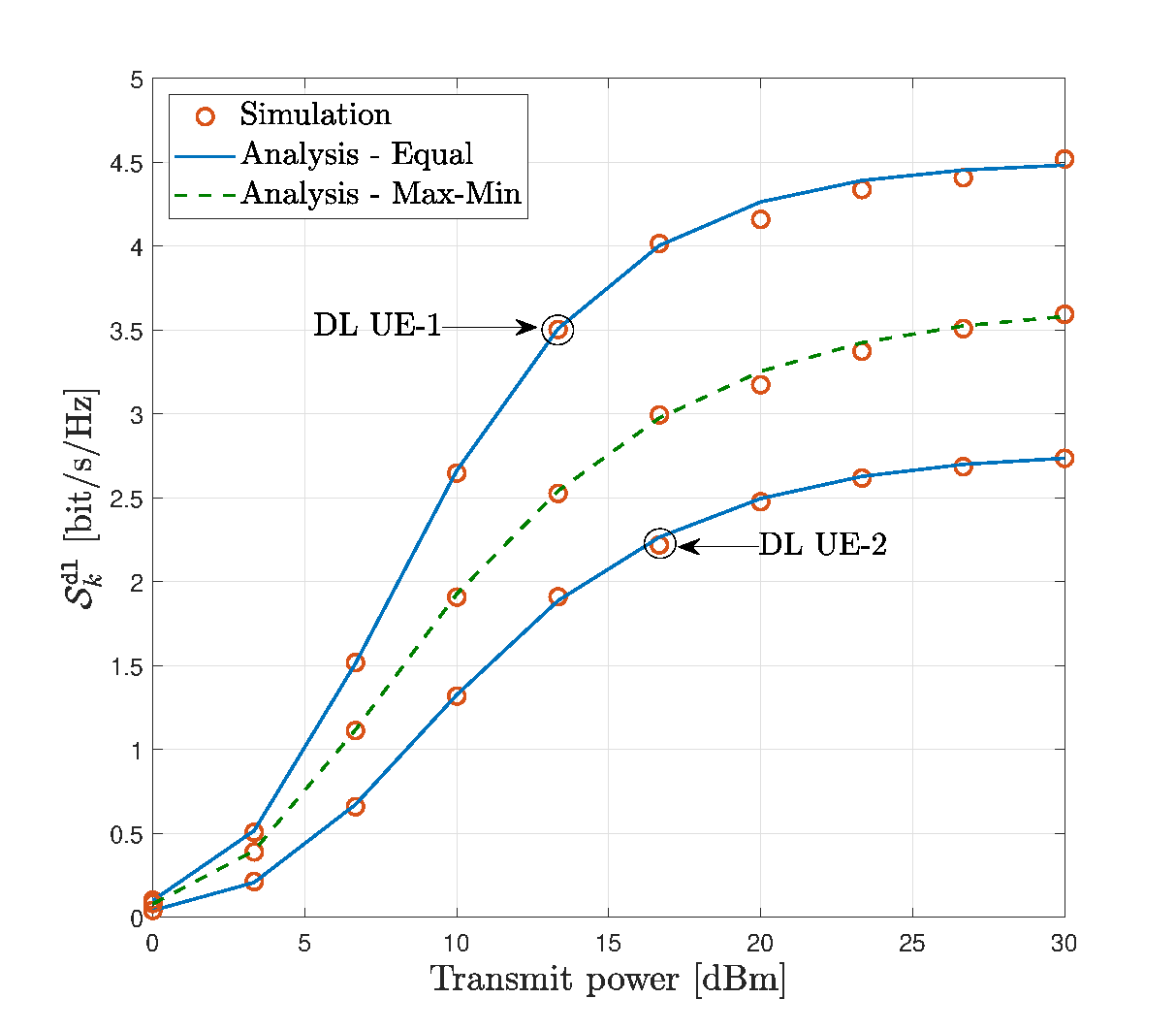}
    \caption{DL SE versus the transmit power for $M=16$, $N_t=N_r=4$, $K=L=2$, and $\tau_p=K+L$.} 
    \label{DLRate_ptx} \vspace{-1em}
\end{figure}

\begin{figure}[!t]\vspace{-0mm}	
    \centering
    \includegraphics[width=0.48\textwidth]{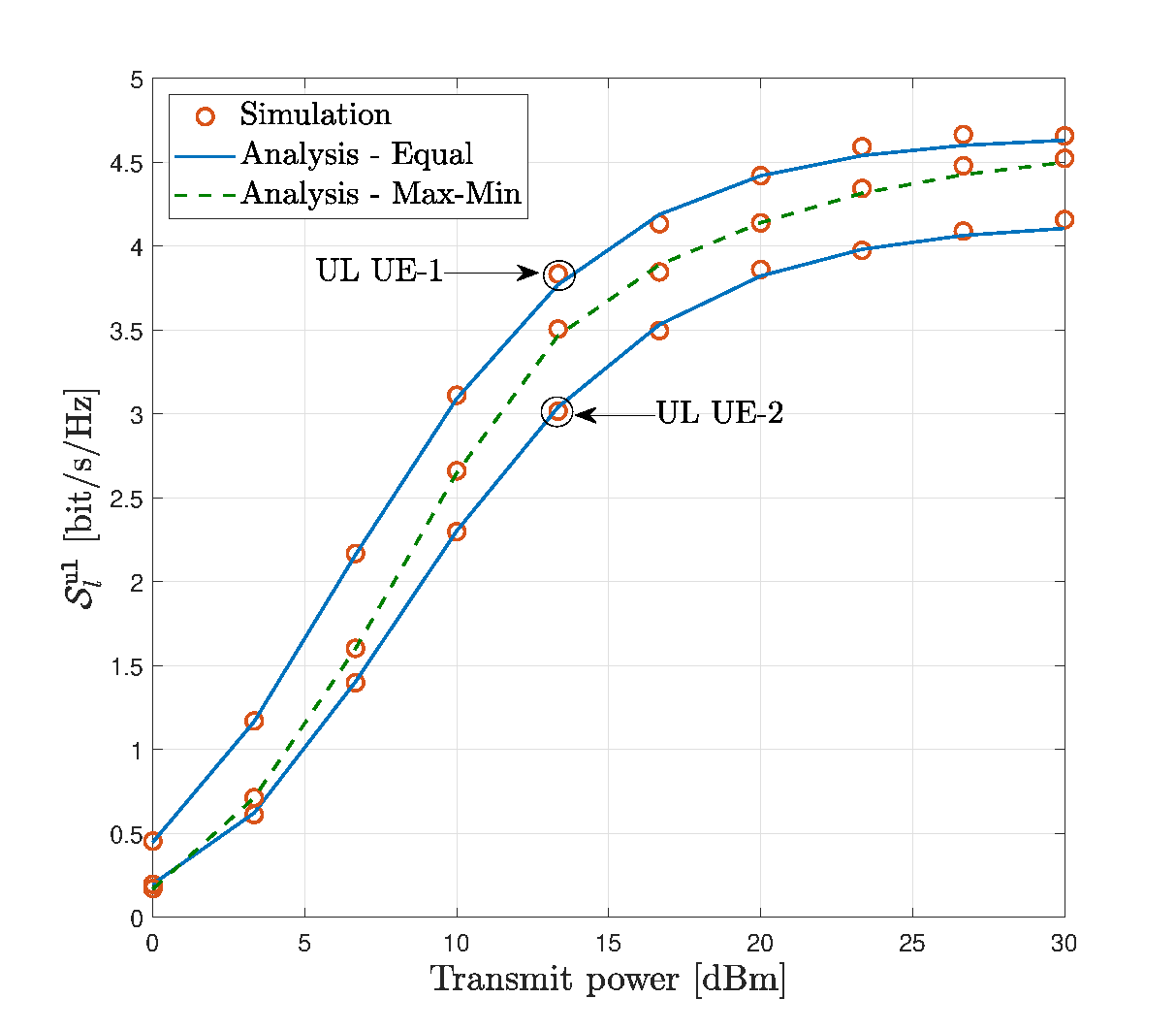}
    \caption{UL SE versus the transmit power for $M=16$, $N_t=N_r=4$, $K=L=2$, and $\tau_p=K+L$.} 
    \label{ULRate_ptx} \vspace{-1em}
\end{figure}

Figures~\ref{DLRate_ptx} and~\ref{ULRate_ptx} plot the DL SE and the UL SE, respectively, as functions of the respective transmit powers, i.e., per-AP DL transmit power, $p_d$, and per-user UL transmit power, $p_u$. The two figures reveal that when equal power distribution is used at APs and UL users, the DL and UL users experience distinct rates. Clearly, the near-far effect caused by the spatial distribution of users and APs has a detrimental impact on the user SEs. However, with the proposed max-min power control, all the DL and UL users, regardless of their location, attain their respective common rates. The proposed scheme thus can guarantee uniform QoS for all user nodes. 

\begin{figure}[!t]\vspace{-0mm}	
    \centering
    \vspace{-4.3em}
    \includegraphics[width=0.5\textwidth]{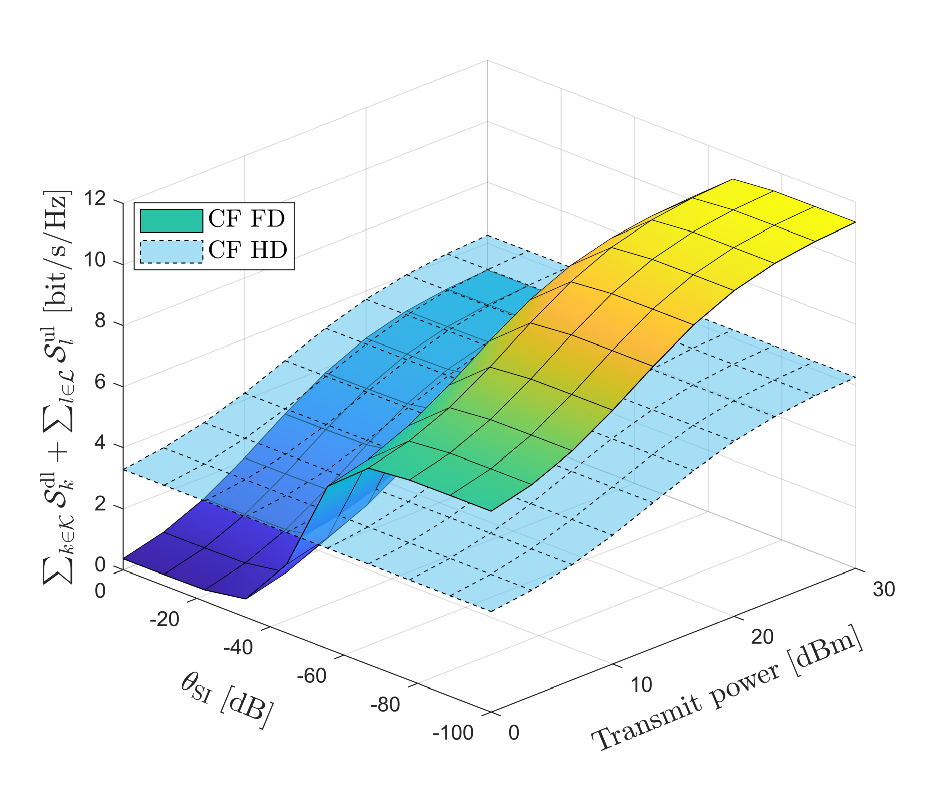}
    \caption{A comparison of sum SE of FD and HD CF mMIMO with transmit power allocation for $M=16$, $N_t=N_r=4$, $K=L=2$, and $\tau_p=K+L$.}
    \label{FD_HD_Comp} \vspace{-1em}
\end{figure}

Figure~\ref{FD_HD_Comp} depicts a comparison of the sum SE for FD and HD CF mMIMO as a function of the transmit power ($p_d=p_u$) and SI suppression factor ($\theta_{\rm{SI}}$). The HD CF system operates in TDD mode to facilitate the UL and DL transmission, eliminating the requirement for SI cancellation compared to the FD. Thus, HD CF SE is independent of the SI suppression factor. The figure reveals that the FD system has a significantly higher cumulative SE than an equivalent HD system, given that the SI suppression is sufficient, i.e., $\theta_{\rm{SI}}\leq \qty{-40}{\dB}$. On the other hand, with weak SI suppression, $\theta_{\rm{SI}}\geq \qty{-40}{\dB}$, the SE benefits obtained by the CF FD system through the FD transmissions completely vanish. Furthermore, despite substantial RI suppression, i.e., $\theta_{\rm{SI}}\leq \qty{-40}{\dB}$, the FD sum SE does not double compared to the HD as predicted by theoretical analysis. This is primarily due to the UL-DL interference encountered by the DL user in the CF FD system, which cannot be mitigated by SI suppression at the APs.

\section{Network-Assisted Cell-Free Full-Duplex Communication}~\label{sec:NAFD}
\begin{figure}[t]\centering \vspace{2em}
    \def\svgwidth{260pt} 
    \fontsize{8}{8}\selectfont 
    \graphicspath{{Figures/}}    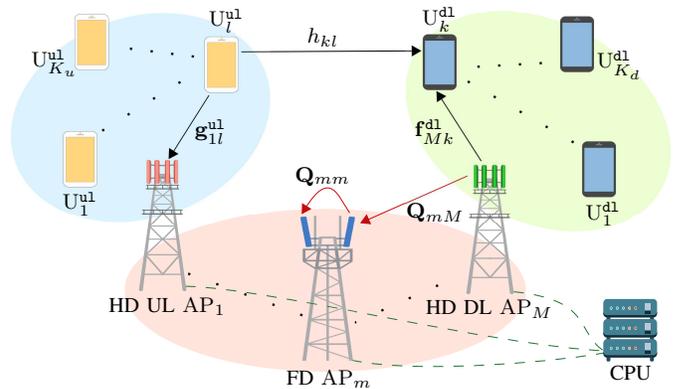 \vspace{0mm}
    \caption{NAFD CF  system with HD and FD APs.}\vspace{-1em}
    \label{fig:NAFD_CFmMIMO}
\end{figure}

While CF FD networks support simultaneous UL and DL transmissions over the same time-frequency resources, they suffer from increased interference in both directions due to concurrent transmissions. Additionally, equipping a large number of APs with FD transceivers imposes significant costs on service providers and increases the overall power consumption of the network. The increase in the power consumption is driven by the energy-intensive SI cancellation circuits at the APs, as well as the power demands on the fronthaul links, since all APs simultaneously use these links to transmit and receive data payloads from the CPU. The key concept behind NAFD CF  is to virtually realize the functionality of a CF FD network by using HD APs~\cite{Wang2020}. This approach offers an opportunity to efficiently manage interference and reduce the high energy consumption associated with FD transceivers and fronthaul links, while still capitalizing on the benefits of simultaneous UL and DL transmissions across the network, as shown in Fig.~\ref{fig:NAFD_CFmMIMO}. 

A NAFD architecture shows similarities to established topologies, such as dynamic-TDD~\cite{Haas:JSAC:2001,Liu:TCOm:2017,Jiamin:TWC:2018}, dynamic UL-DL configuration in time-division Long Term Evolution~\cite{Shen:MCOM:2012}, coordinated multipoint for wireless IBFD (CoMPflex)~\cite{Thomsen:WCL:2016}, and bidirectional dynamic networks~\cite{Xin:ACCESS:2017},  which were proposed to manage the asymmetric UL-DL data traffic in wireless (cellular) networks. In dynamic TDD, a particular time slot can be assigned dynamically for reception or transmission based on the UL-DL traffic demands in each cell. However, this approach does not completely address the diverse data needs within the cells, as UEs assigned a DL (UL) time slot in a cell must wait for the UL (DL) time slot to transmit their data. Moreover, the effectiveness of the dynamic TDD is constrained by inter-BS interference and the need for strict synchronization among cells, which necessitates cell cooperation, causing significant overhead and complexity. The reasoning for implementing the dynamic UL-DL configuration in time-division Long Term Evolution is to ensure that different small cells within diverse cellular networks can support transmission in opposite directions~\cite{Shen:MCOM:2012}. CoMPflex, which is motivated by the coordinated multipoint concept in cellular networks, involves emulating a FD BS and using two spatially separated and coordinated HD BSs~\cite{Thomsen:WCL:2016}. In bidirectional dynamic networks, the number of UL and DL remote radio heads is adjusted flexibly to facilitate simultaneous UL and DL communications~\cite{Xin:ACCESS:2017}.

In this context, NAFD systems offer several advantages over the conventional CF FD counterparts:
\begin{itemize}
    \item The NAFD framework eliminates the SI within the APs, simplifying the transceiver design and avoiding the need for power-hungry SI cancellation circuits.
    \item   Some APs will handle UL transmission, while others will manage DL transmission, significantly reducing the fronthaul data traffic and associated energy consumption. 
    \item  The flexible adjustment of the UL and DL modes at the HD APs can efficiently cope with the inter-AP interference (a main bottleneck in CF FD) and, at the same time, can accommodate the increasing time-varying traffic asymmetry between the UL transmission and DL reception for 6G wireless. 
\end{itemize}
These benefits have driven the application of the NAFD concept in CF  systems to enable simultaneous UL and DL communications on the same frequency band.


\subsection{Performance Analysis}
The concept of NAFD CF  was first introduced in~\cite{Wang2020}, where HD APs were used to support both UL and DL transmissions. Specifically, the SE of a NAFD CF  network was analyzed for imperfect CSI and spatial correlations. By applying large-dimensional random matrix theory, deterministic equivalents for the sum rate were derived for both UL, using an MMSE receiver, and DL, employing regularized zero-forcing (RZF) and zero-forcing (ZF) precoders. The authors identified inter-AP interference (between DL and UL APs) and inter-user interference (UL-to-DL interference) as the primary bottlenecks in system design. In a static scenario, where the APs have fixed operational modes (either UL or DL), they assumed that inter-AP DL-to-UL channels could be estimated using pilot signals with minimal overhead due to the quasi-static nature of the links. To mitigate UL-to-DL interference, they proposed a user scheduling scheme. 

In~\cite{Wang2020}, a fixed network configuration for APs was conceived, where certain APs are designated for UL transmission and others for DL transmission. However, in dynamic scenarios with time-varying traffic asymmetry between DL transmission and UL reception, static configurations limit the full potential of the NAFD concept. In these cases, dynamically adjusting AP operation modes to balance UL and DL transmissions would better harness  NAFD's advantages. To address this shortcoming, the authors in~\cite{Mohammadali:JSAC:2023} developed a comprehensive performance analysis framework under dynamic AP mode selection, offering both SE and EE analysis. They derived closed-form expressions for the UL and DL SE, assuming MRC and MRT processing at the APs and DL UEs, while accounting for large-scale fading decoding (LSFD) coefficients in the UL. Additionally, they introduced a total power consumption model that incorporates the fronthaul power consumption.

\subsection{Channel Estimation}
To fully leverage the benefits of NAFD CF  systems, accurate CSI is essential at both the CPU and terminals, to address the following aspects:
\begin{itemize}
    \item The CPU requires  CSI between the APs and terminals to design beamforming and receivers. This CSI can be obtained through pilot training under the TDD protocol, similar to conventional FD and HD CF  networks.
    \item In NAFD distributed massive MIMO systems, the simultaneous DL transmission and UL reception cause inter-AP interference, as the DL transmissions interfere with the UL reception. This interference remains the primary factor limiting UL SE. Therefore, effective interference cancellation is essential for enhancing the UL SE in NAFD distributed mMIMO systems. To this end, the estimation of the interference channel matrix between the DL-APs and UL APs is necessary.
\end{itemize}
Although the two CSI requirements mentioned are similar to those in conventional CF FD networks, there is a key distinction in NAFD systems from a channel estimation perspective. In NAFD CF  systems, each terminal is usually served by a subset of nearby APs, weakening the channel hardening effect. Consequently, the DL terminals must estimate the CSI between themselves and the DL-APs directly, rather than depending on statistical CSI.

Li~\ettall~\cite{Jiamin:TWC:2021} proposed a method to estimate the effective CSI—specifically, the inner products of beamforming and channel vectors—through a beamforming training scheme. This approach allows the CPU to estimate the effective CSI between the DL APs and UL APs, while the DL users can estimate the effective CSI between DL APs and themselves. Under this approach, they derived closed-form expressions for the UL and DL achievable rates with various receivers and beamforming strategies. Based on these expressions, they developed an efficient power allocation scheme that depends only on slowly varying large-scale fading, framed within a multi-objective optimization perspective. Note that the channel estimation proposed in~\cite{Jiamin:TWC:2021} applies to systems with fixed AP modes. However, in dynamic NAFD systems, where the AP operation mode is designed based on UL and DL traffic, the inter-AP DL-to-UL channel estimation method in~\cite{Wang2020} becomes infeasible.

\subsection{Resource Allocation}
Selecting the operation mode of APs (i.e., UL or DL) introduces an additional degree of freedom in NAFD systems compared to their FD counterparts. As a result, beyond the design parameters in CF FD systems—such as power control at the APs and UL users, and LSFD design at the CPU—the operation mode of the APs can also be optimized to meet the network requirements. Recent studies revealed that inter-AP interference becomes more manageable under a flexible UL/DL operation mode assignment at the HD APs. Zhu~\ettall~\cite{Zhu2021Optimization} proposed a duplex mode selection algorithm based on parallel successive convex approximation, along with a reinforcement learning algorithm based on Q-learning, aiming at maximizing the user sum rate. Chowdhury~\ettall~\cite{Chowdhury:TCOM:2022} formulated the AP-scheduling problem as one of maximizing the sum UL-DL SE, by considering MRT/RZF precoding in the DL and MRC/MMSE in the UL based on the locally estimated channels. They proposed a greedy algorithm for dynamic AP-scheduling, where, at each step, the transmission mode of the AP that maximizes the incremental SE is added to the already scheduled AP-subset. Moreover, an iterative pilot allocation algorithm, based on locally
estimated channel statistics at the APs, was developed. Sun~\ettall~\cite{Sun:JSYS:2024} proposed a “preallocation—optimization” mechanism for AP duplex mode optimization, where the preallocation part and the optimization part can help NAFD CF  networks effectively handle the time-varying active UEs and network load in massive Internet of Things (IoT) scenarios.  The preallocation part involves a network load prediction algorithm based on an autoregressive integrated moving average model, ensuring accurate load forecasting and efficient preallocation of resource blocks. In the optimization part,  deep reinforcement learning-based and  hierarchical reinforcement learning-based AP duplex mode optimization algorithm were developed to solve the multi-objective optimization problem of AP mode optimization. While the proposed algorithms offer improved performance compared to fixed duplex mode networks, some users' UL and DL SE is still compromised due to inefficient resource allocation.

The flexibility of NAFD design has been leveraged for secure transmission. Specifically, Xia \textit{et al.} \cite{Xia:JSYS:2023} focused on the joint design of transmit information vectors, artificial noise covariance matrices, AP mode selection vectors, and receiver vectors. Their goal was to maximize the secure DL and UL transmission rates while adhering to the sum transmission power constraints at the APs and UL users.

Another line of research has integrated QoS requirement constraints for UL and DL users into the AP mode assignment problem, making the associated optimization challenge more complex~\cite{Xia:TWC:2021, Xia:TVT:2023, Zhu:Access:2022,  Mohammadali:JSAC:2023}. More specifically,  Xia~\ettall~\cite{Xia:TWC:2021} considered the SE maximization problem in  NAFD systems, where the APs operate in either FD mode or HD mode. By considering the QoS requirements of both the UL and DL users, along with fronthaul capacity constraints, they investigated the joint user selection and transceiver design. Xia~\ettall~\cite{Xia:TVT:2023} addressed the EE maximization problem for a CF  system with NAFD, considering QoS requirements, transmit power constraints, fronthaul capacity constraints and energy harvesting constraints. In~\cite{Xia:TWC:2021, Xia:TVT:2023}, the operation mode of the APs is fixed, leaving the potential of dynamic network configuration unexplored.

Zhu~\ettall~\cite{Zhu:Access:2022} proposed a load-aware dynamic mode selection scheme for the APs, aiming to maximize the UL-DL sum-rate of the network while considering the per-user traffic load. They investigated centralized Q-learning and distributed multi-agent Q-learning algorithms with varying complexities, demonstrating that the former is more suitable for real-world applications due to its smaller storage unit and lower complexity.  The proposed algorithms in~\cite{Xia:TWC:2021,Zhu:Access:2022,  Xia:TVT:2023} rely on instantaneous CSI at the APs and CPU, thereby imposing high overhead and complexity on the system. To tackle this issue, Mohammadi~\ettall~\cite{Mohammadali:JSAC:2023} proposed a joint AP mode assignment, power control, and LSFD design algorithm to maximize the SE and EE. Their results in \cite{Mohammadali:JSAC:2023} confirmed that by optimizing the operation mode at the APs, the EE of NAFD CF  systems achieves up to a two-fold improvement over CF FD systems. Vu~\ettall~\cite{Tung:EUSIPCO:2023} proposed a multipair decode-and-forward NAFD CF  relaying system, where the APs with a DL mode serve destination nodes and those with a UL mode serve source nodes, at the same time. A joint optimization approach to design the AP mode assignment, power control, and LSFD weights was proposed, which maximizes the sum SE of transmission pairs, under per-pair SE requirements.

Building on the NAFD CF  concept, AP mode selection has been incorporated into SWIPT-enabled CF  systems~\cite{Mohammadi:GC:2023} and CF  surveillance networks~\cite{Mobini:IoT:2024}. In these scenarios, two distinct groups of APs are designated to carry out separate functions simultaneously over the same time-frequency resources.

\begin{table*}
	\centering
	\caption{Summary of NAFD CF  literature.}\label{tabel:TWRN}
	\small
\begin{tabular}
{|C{2cm}|c|c|c|c|c|c|c|C{7cm}|}
	\hline
	 \multirow{2}{*}{\begin{tabular}[c]{@{}c@{}}\textbf{Category}\end{tabular}} & \multirow{2}{*}{\textbf{Lit.}} & \multicolumn{3}{c|}{\textbf{AP}} & 
	\multicolumn{3}{c|}{\textbf{UE}} & \multirow{2}{*}{\textbf{Technical Contribution}}\\

	\cline{3-8}
	\centering
	& &\textbf{FD}  &\textbf{UL} &\textbf{DL}  &\textbf{FD} &\textbf{UL} &\textbf{DL}  & \\
    \hline \hline

	\multirow{3}{*}{\begin{tabular}[c]{@{}c@{}}Performance \\Analysis\end{tabular}} & ~\cite{Wang2020} &-   &\checkmark &\checkmark &- &\checkmark &\checkmark & Sum-rate analysis for UL with MMSE receiver and DL with RZF and ZF precoders under fixed AP mode of operation.\\
   \cline{2-9}
   
     & ~\cite{Mohammadali:JSAC:2023} &-  &\checkmark &\checkmark &- &\checkmark &\checkmark & SE and EE with MRC/MRT processing under flexible AP modes of operations.  \\

	\hline

   Channel Estimation
    & ~\cite{Jiamin:TWC:2021} &\checkmark   &\checkmark &\checkmark &- &\checkmark &\checkmark &Beamforming training scheme to estimate the effective
     CSI between the DL APs and UL APs at CPU as well as effective
     CSI between DL APs and DL users. \\
	\hline

 \multirow{22}{*}{\begin{tabular}[c]{@{}c@{}}Resource \\Allocation\end{tabular}}

    & ~\cite{Chowdhury:TCOM:2022} &-   &\checkmark &\checkmark &- &\checkmark &\checkmark & Development of a greedy-based AP mode selection algorithm for sum UL-DL SE maximization.\\
	\cline{2-9}
 
    & ~\cite{Zhu2021Optimization} &-   &\checkmark &\checkmark &- &\checkmark &\checkmark & AP duplex mode selection for sum-rate maximization with successive convex approximation and a reinforcement learning algorithm.  \\
	\cline{2-9}
 
    & ~\cite{Sun:JSYS:2024} &-   &\checkmark &\checkmark &- &\checkmark &\checkmark & AP duplex mode optimization, aiming to address the dynamic massive IoT scenarios and achieve a balance between SE and resource utilization.\\
	\cline{2-9}

     & ~\cite{Xia:JSYS:2023} &\checkmark   &\checkmark &\checkmark &- &\checkmark &\checkmark & Joint AP duplex mode and transceivers optimizing to maximize the overall secrecy SE with artificial noise transmission.\\
	\cline{2-9}

     & ~\cite{Xia:TWC:2021} &\checkmark   &\checkmark &\checkmark &- &\checkmark &\checkmark & Maximizing the SE and number of serving UEs under the QoS requirements of UL and DL UEs, with fixed AP mode of operation.\\
	\cline{2-9}

     & ~\cite{Xia:TVT:2023} &\checkmark   &\checkmark &\checkmark &- &\checkmark &\checkmark & Joint strategy of EH, fronthaul compression, and CLI cancelation to maximize the EE under fixed AP mode of operation. \\
	\cline{2-9}
 
    & ~\cite{Zhu:Access:2022} &-   &\checkmark &\checkmark &- &\checkmark &\checkmark & Load-aware dynamic mode selection scheme of APs based on centralized Q-learning and distributed multi-agent Q-learning method. \\
	\cline{2-9}

    & ~\cite{Mohammadali:JSAC:2023} &-  &\checkmark &\checkmark &- &\checkmark &\checkmark & Joint AP mode selection, power control, and LSFD design to maximize SE and EE. \\
	\cline{2-9}

    & ~\cite{Tung:EUSIPCO:2023} &-  &\checkmark &\checkmark &- &\checkmark &\checkmark & Multipair decode-and-forward NAFD CF  relaying system. \\
    \hline


\end{tabular}
\vspace{-1em}
\end{table*}

\textbf{Case study and discussion:} Consider a NAFD CF  system, where $M$ APs serve $L$ UL UEs and $K$ DL UEs -- Fig.~\ref{fig:NAFD_CFmMIMO}. Define the sets $\mathcal{M}\triangleq\{1,\ldots,M\}$, $\K\triangleq \{1,\dots,K\}$ and  $\mathcal{L}\triangleq\{1,\ldots,L\}$ as collections of indices of the APs, DL UEs, and UL UEs, respectively.
Each AP  connects to the CPU via a high-capacity fronthaul link. The DL and UL transmissions are in the same time-frequency block via HD (and IBFD) APs.  Each UE has one single antenna, while each AP is equipped with  $N$ transmit and $N$ receive RF chains. The HD APs are dynamically assigned as DL APs or UL APs to accommodate the DL and UL transmissions, respectively.

The channel coefficient vector between the $k$-th DL UE, $\Ukdl$, ($l$-th UL UE, $\Ulul$) and the $m$-th AP, $\gmkd\in\mathbb{C}^{N \times 1}$ ($\gmlu\in\mathbb{C}^{N \times 1}$), $\forall k \in \K$, $l \in \mathcal{L}$, $ m \in \MM $, is modeled as $\gmkd=\sqrt{\betamkd}\tgmkd,~(\gmlu=\sqrt{\betamlu}\tgmlu) $, where $\betamkd$ ($\betamlu$) is the large-scale fading coefficient and $\tgmkd\in\mathbb{C}^{N \times 1}$ ($\tgmlu\in\mathbb{C}^{N \times 1}$) is the small-scale fading vector whose elements are i.i.d. $\mathcal{CN} (0, 1)$ random variables (RVs). The channel gain between the UL UE $l$ to the DL UE $k$ is denoted by $h_{kl}$. It can be modeled as $h_{kl}=(\betakldu)^{1/2}\tilde{h}_{kl}$, where $\betakldu$ is the large-scale fading coefficient and $\tilde{h}_{kl}$ is a $\mathcal{CN}(0,1)$ RV. Finally, the interference links among the APs are modeled as Rayleigh fading channels. Let $\qQ_{mi}\in \mathbb{C}^{N\times N}$, $i\neq m$, be the channel matrix from AP $m$ to AP $i$, $\forall m,i\in\MM$, whose elements are  i.i.d. $\mathcal{CN}(0,\betami)$ RVs. Here, $\qQ_{mm}, \forall m$ denotes the SI channel at the IBFD APs, whose elements are i.i.d. $\mathcal{CN}(0,\zeta_{Q_{mm}}\triangleq \theta_{\rm{SI}})$ RVs.

\textit{DL Data Transmission:} The  signal received at $\Ukdl$ is 
\begin{align}~\label{eq:ykdl}
y_k^{\dl}
&=
\sqrt{\rho_d}
\sum\nolimits_{m \in \mathcal{M}} a_m\mu_{mk}
\left(\gmkd\right)^\dag\wmkdl
s_{k}^{\dl}
\nonumber\\
&+
\sqrt{\rho_d}
\sum\nolimits_{m \in \mathcal{M}}
\sum\nolimits_{k'\in\K \setminus k} 
a_m\mu_{mk'}
\left(\gmkd\right)^\dag \wmkpdl
s_{k'}^{\dl}
\nonumber\\
&+
\sum\nolimits_{l\in \mathcal{L}}h_{k\ell}x_{\ell}^{\ul}+w_{k}^{\dl},
\end{align}
where $\rho_d$ denotes the normalized DL SNR, $a_m$ denotes the binary AP mode assignment variable with $a_m=1$ indicating that AP $m$ operates in the DL mode; $\mu_{mk}$ denotes the power control coefficient at AP $m$ corresponding to $\Ukdl$, $\wmkdl\in\mathbb{C}^{N\times 1}$ is the precoding vector at AP $m$ for DL UE $k$,  $s_k^\dl\sim\mathcal{CN}(0,1)$ denotes the intended symbol for $\Ukdl$, $w_{k}^{\dl}\sim\mathcal{CN}(0,1)$ is the AWGN at DL UE $k$. Note that the third term in~\eqref{eq:ykdl} is the cross-link interference (CLI) caused by the UL UEs due to concurrent transmissions of DL and UL UEs over the same frequency band, while $x_\ell^\ul$ denotes the transmit signal from the $\Ulul$.
We assume that AP $m$ uses the available CSI to locally form the MRT precoding vector for DL UE $k \in \Wm$ as  $\wmkdlmr = \hgmkd$. 

With MRT precoding, the achievable DL SE at the $k$-th DL UE is given by 
	\begin{align}~\label{eq:DL:SE}
&\mathcal{S}_{k}^\dl (\qa, \boldsymbol \mu, {\boldsymbol{\varsigma}}) =  \nonumber\\
&\hspace{1em}\frac{\tau_c-\tau_p}{\tau_c}
\log_2 \left(1  
    + \frac{\rho_d(\sum\nolimits_{m \in \MM} N  a_m\mu_{mk} \gamdmk)^2}{\Omega_k(\boldsymbol \mu, {\boldsymbol{\varsigma}}, \qa)}
	\right),
	\end{align}
where 
\begin{align}
     \Omega_k (\boldsymbol \mu, {\boldsymbol{\varsigma}}, \qa) &=\!
    {\rho_d}\!\!
    \sum\nolimits_{m \in \MM} 
    \sum\nolimits_{k'\in\mathcal{K}} 
    N  a_m\mu_{mk'}^2
\gamdmkp\betamkd 
\nonumber\\
&
\!+\rho_u\sum\nolimits_{l\in\mathcal{L}}{ \tilde{\varsigma}_l}\betakldu\!+\!1.
\end{align}    

Moreover, $\qa \triangleq \{a_m\}$; $\boldsymbol{\mu}\triangleq \{\mu_{mk}\}$ denote the DL power control coefficients at AP $m\in\MM$ towards $\Ukdl$, $\forall k\in\K$; ${\boldsymbol{\varsigma}} \triangleq \{{\varsigma}_{l}\}$, present the UL power control coefficients for $\Ulul,$ $\forall l\in\mathcal{L}$;  $\rho_u$ denotes the normalized UL SNR; $\gamdmk$ is the variance of the MMSE estimate of $\gmkd$, i.e., $\hgmkd\sim \mathcal{CN}(\boldsymbol{0}, \gamdmk\qI_N)$.

\textit{UL Data Transmission:} The transmit signal  from  UL UE $\ell$ is represented by $x_{\ell}^\ul  = \sqrt{\rho_u {\varsigma}_\ell} s_{\ell}^{\ul}$, 
where $s_{\ell}^\ul$ (with $\Ex\left\{|s_{\ell}^\ul|^2\right\}=1$) and $\rho_u$ denote the transmitted symbol by $\Ulul$ and the normalized transmit power at each UL UE, while ${\varsigma}_{\ell}$ is the transmit power control coefficient at $\Ulul$ with $ 0\leq {\varsigma}_{\ell} \leq 1, \forall \ell$.

The received signal $\qy_{m}^{\ul}\in\mathbb{C}^{N \times 1}$ at AP $m$ in the UL mode can be written as
\begin{align}\label{eq:ymul}
\qy_{m}^{\ul}
&=
\sqrt{\rho_u}\sum\nolimits_{l \in \mathcal{L}}\sqrt{b_m \tilde{\varsigma}_{l}}\qg_{ml}^{\ul} s_{l}^{\ul}
\nonumber\\
&
+
\underbrace{\sqrt{\rho_d}\sum\nolimits_{i\in\mathcal{M}\setminus m}\sum\nolimits_{k\in \K}
\sqrt{b_m a_i } \mu_{ik}
\qQ_{mi}
\wikdl s_k^\dl}_{\text{Intra-AP interference}}\nonumber\\
&
\hspace{0em}
+
\underbrace{\sqrt{\rho_d}\sum\nolimits_{k\in \K}
\sqrt{b_m a_m } \mu_{mk}
\qQ_{mm}
\wmkdl s_k^\dl
}_{\text{Inter-AP (SI) interference}}+\sqrt{b_m}\qw_{m}^{\ul},
\end{align}
where $b_m$ denotes the binary AP mode assignment variable with $b_m=1$ indicating that AP $m$ operates in the UL mode and $\qw_{m}^{\ul}$ is the AWGN vector with $\mathcal{CN}(0,1)$ distributed elements. In~\eqref{eq:ymul}, the second term captures the interference from APs transmitting towards DL UEs and the third term represents the SI term wherever AP $m$ is FD-enabled (i.e., $b_m a_m=1$).

In order to improve the achievable UL SE, the signals forwarded to the CPU from the UL APs $m$, corresponding to UE $l$, are further multiplied by the LSFD weight $\alpha_{ml}$, $\forall m, l$.   The aggregated received signal for UL UE $\ell, \forall \ell$ with $|\alphml|^2\leq 1, \forall \ell, m$ at the CPU can be written as
\begin{align}\label{eq:rul}	\qr_{\ell}^{\ul}=\sum\nolimits_{m\in\MM}\alphml(\wmlul)^\dag\qy_{m}^{\ul}. 
\end{align}
Finally, $s_{\ell}^{\ul}$ is detected from $\qr_{\ell}^{\ul}$. 

We assume that MRC is performed at the IBFD/UL HD AP with $\wmlulmr=\hgmlu$ and data is transferred to CPU for UL data reception. The achievable UL SE for $l$-th UL UE at the CPU  is given by
\begin{align}
    \label{eq:UL:SE}
	\mathcal{S}_{l}^\ul (\qa,\qb, \boldsymbol{\varsigma}, \boldsymbol{\mu}, \boldsymbol{\alpha} )
	\!= \!\frac{\tau_c\!-\!\tau_p}{\tau_c}\log_2
	(1\!+\! \SINR_{l}^\ul(\qa, \qb, \boldsymbol{\varsigma}, \boldsymbol{\mu}, \boldsymbol{\alpha} )),
\end{align}
where $\qb \triangleq \{b_m\}$  and
\begin{align}\label{eq:SINRMA_ZF1}
&\SINR_{l}^\ul( \qb, \boldsymbol{\varsigma}, \boldsymbol{\mu}, \boldsymbol{\alpha})=
\nonumber\\
&\hspace{1em}\frac{
	\rho_{u} \Big(
 {\sum_{\substack{m\in\MM}}}  N
 \alphml \sqrt{b_m \varsigma_{l}} {\gamuml}\Big)^2
	}
	{\Phi_{l}(\qb, \boldsymbol{\varsigma}, \boldsymbol{\mu}, \boldsymbol{\alpha})+
		\rho_d 
		\sum_{\substack{i\in\MM}}
		\sum_{q\in\K}
		 a_{i}\mu_{iq}^2 \gamma_{iq}^{\dl}\Psi_{l i}( \qb, \boldsymbol{\alpha})
},
\end{align}
where $\gamuml$ is the variance of the MMSE estimate of $\gmlu$ and
\begin{subequations}~\label{eq:vhomli:mumlpzf}
\begin{align}
\Phi_{l}(\qb, \boldsymbol{\varsigma}, \boldsymbol{\mu}, \boldsymbol{\alpha})&=
  \rho_{u}\sum\nolimits_{m\in\MM}
		 \sum\nolimits_{l'\in\mathcal{L}}\!   
        \Big(\Ntx
     \alphml^2 b_m \varsigma_{l'}
	{\gamuml\zeta_{g_{ml'}}^{\ul}}
  \nonumber\\
  &\hspace{2.5em}
  +  \Ntx
 		\alphml^2b_m\gamuml\Big),
  \\
 \Psi_{l i}(\qb, \boldsymbol{\alpha})&\!= \!
 \sum\nolimits_{m\in\MM}
 N^2
 \alphml^2 b_m
 {\gamuml}
      \betami.~\label{eq:vho:mli:pzf}
\end{align}    
\end{subequations}

\textit{Power consumption:} Let $P_{\ell}$ and $P_{\mathtt{U},\ell}$ be the power consumption for transmitting signals and the required power consumption to run the circuit components for the UL transmission at $\Ulul$.  Moreover,  assume that $P_{\mathtt{D},k}$ denotes the power consumption to run the circuit components for the DL transmission at $\Ukdl$; $\Pbhm$ is the power consumed by the fronthaul link between the CPU and AP $m$. Therefore, the total power consumption over the considered NAFD CF  system is modeled as~\cite{Emil:TWC:2015:EE,Ngo2018}
\begin{align}~\label{eq:Ptotal}
P_\mathtt{total} &= \sum\nolimits_{\ell\in\mathcal{L}} (P_{\ell} + P_{\mathtt{U},\ell}) + \sum\nolimits_{m\in\mathcal{M}} P_m  \nonumber\\
&\hspace{2em}+ \sum\nolimits_{k\in\K} P_{\mathtt{D},k} + \sum\nolimits_{m\in\mathcal{M}} \Pbhm,
\end{align}
where $P_m$ denotes the power consumption at AP $m$ that includes the power consumption of the transceiver chains and the power consumed for the DL or UL transmission. The power consumption $P_m$ can be modeled as~\cite{Mohammadali:JSAC:2023}
\begin{align}~\label{eq:Pm}
\nonumber
&P_m (\boldsymbol a, \boldsymbol b,\boldsymbol \theta) = 
\frac{1}{\zeta_m}\rho_{d}\Sn\left(N\sum\nolimits_{k\in\mathcal{K}} \gamdmk \mu_{mk} \right) 
+ a_m N P_{\mathtt{cdl},m} \nonumber\\
&\hspace{4em}
+ b_m N P_{\mathtt{cul},m},
\end{align}
where $0<\zeta_m\leq 1 $ is the power amplifier efficiency at the $m$-th AP, $\Sn$ is the noise power; $P_{\mathtt{cdl},m}$ and $P_{\mathtt{cul},m}$ is the internal power required to run the circuit components (e.g., converters, mixers, and filters) related to each antenna of AP $m$ for the DL and UL transmissions, respectively. The power consumption at $\Ulul$ is given by
\begin{align}~\label{eq:Pl}
P_{\ell}= \frac{1}{\chi}\rho_u\Sn {\varsigma}_{\ell}, 
\end{align}
where $\chi$ is the power amplifier efficiency at UL UEs.

The power consumption of the fronthaul signal load to each AP $m$ is proportional to the fronthaul rate as \cite{Ngo2018}
\begin{align}~\label{eq:Pbhm}
\Pbhm = a_m P_{\mathtt{fdl},m} + b_m P_{\mathtt{ful},m} + R_m
P_{\mathtt{bt},m},
\end{align}
 where $P_{\mathtt{bt},m}$ is the traffic-dependent fronthaul power (in Watt per bit/s) and $R_m$ is the fronthaul rate between AP $m$ and the CPU, given by
\begin{align}~\label{eq:Sx}
R_m &=  B \Big( a_m \sum\nolimits_{\ell\in\mathcal{L}} \mathcal{S}_{\ell}^{\ul} (\qb, \boldsymbol \varsigma, \boldsymbol \mu, \boldsymbol \alpha)   \nonumber\\
&\hspace{4em}
+ b_m \sum\nolimits_{k\in\mathcal{K}}\mathcal{S}_{k}^{\dl} (\qa, \boldsymbol \mu, {\boldsymbol{\varsigma}}) \Big),
\end{align}
By substituting~\eqref{eq:Pm},~\eqref{eq:Pl}, and~\eqref{eq:Pbhm} into~\eqref{eq:Ptotal}, we have
\begin{align}~\label{eq:Ptotal:final}
P_\mathtt{total} (\qa, \qb, \boldsymbol \varsigma, \boldsymbol \mu, \boldsymbol \alpha)
&= 
  \sum_{m\in\mathcal{M}} \frac{N\rho_{d}\Sn}{\zeta_m}\left(\sum\nolimits_{k\in\mathcal{K}} \gamdmk \mu_{mk} 
\right) 
\nonumber\\
&\hspace{-2em}
+\sum\nolimits_{\ell\in\mathcal{L}} \frac{\rho_u\Sn}{\chi} {\varsigma}_{\ell} +\PUfix + R_m
P_{\mathtt{bt},m} \nonumber\\
& \hspace{-2em} 
+  \sum\nolimits_{m\in\mathcal{M}} a_m (N P_{\mathtt{cdl},m} + P_{\mathtt{fdl},m}) 
\nonumber\\
&\hspace{-2em}
+ \sum\nolimits_{m\in\mathcal{M}}  b_m (NP_{\mathtt{cul},m} + P_{\mathtt{ful},m}),
\end{align}
where $\PUfix \triangleq \sum_{k\in\mathcal{L}} P_{\mathtt{U},\ell} +  \sum_{k\in\K} P_{\mathtt{D},k}$. 


\textit{System design:} The AP mode assignment problem involves determining the UL and DL mode assignment vectors $(\qa,\qb)$, the power control coefficients $(\boldsymbol{\varsigma}, \boldsymbol{\mu})$, and the LSFD weight vector $\boldsymbol{\alpha}$. The objective is to maximize the EE of the NAFD CF  while adhering to constraints on the per-user SE, the transmit power at each AP, and the UL UE. More precisely,  the optimization problem  is formulated as
\begin{subequations}\label{P1_prob}
    \begin{eqnarray}
        \mathbf{P2}:
        \underset {\qa,\qb, \boldsymbol{\varsigma}, \boldsymbol{\mu}, \boldsymbol{\alpha}}{\rm{max}} && 
        \mathtt{EE} \triangleq \frac{R_m(\qa,\qb, \boldsymbol{\varsigma}, \boldsymbol{\mu}, \boldsymbol{\alpha})}{P_\mathtt{total}(\qa,\qb, \boldsymbol{\varsigma}, \boldsymbol{\mu}, \boldsymbol{\alpha})}\\
        \text{s.t}\hspace{0.3em} 
        && \mathcal{S}_{\ell}^{\ul} (\qb, \boldsymbol{\varsigma}, \boldsymbol{\mu}, \boldsymbol{\alpha}) \geq \mathcal{S}_\ul^o,~\forall \ell,
		\label{UL:QoS:cons} 
  \\
        &&\mathcal{S}_{k}^{\dl} (\qa, \boldsymbol\mu, {\boldsymbol{\varsigma}}) \geq  \mathcal{S}_\dl^o,~\forall k, 
		\label{DL:QoS:cons}
        \\
        && \sum\nolimits_{k\in\K} \!\! N\gamdmk \theta_{mk}^2 \!\leq\! a_m,
\label{P2_power control}
\\
  && |\alphml|^2\leq 1, \forall \ell, m,   \label{P2_alpha}\\
        && a_m, b_m \in\{0,1\},  \label{P2_AP_mode}\\
        && 0\leq {\varsigma}_{\ell} \leq 1, \forall \ell,
    \end{eqnarray}
\end{subequations}
where $\mathcal{S}_\ul^o$ and $\mathcal{S}_\dl^o$ are the minimum SEs required by the $\Ulul$ and $\Ukdl$, respectively, to guarantee the QoS in the network. Moreover, constraint~\eqref{P2_power control} is the per-AP transmit power constraint.

\begin{figure}[!t]\vspace{-0mm}	
    \centering
     \vspace{-3em}
    \includegraphics[width=0.5\textwidth]{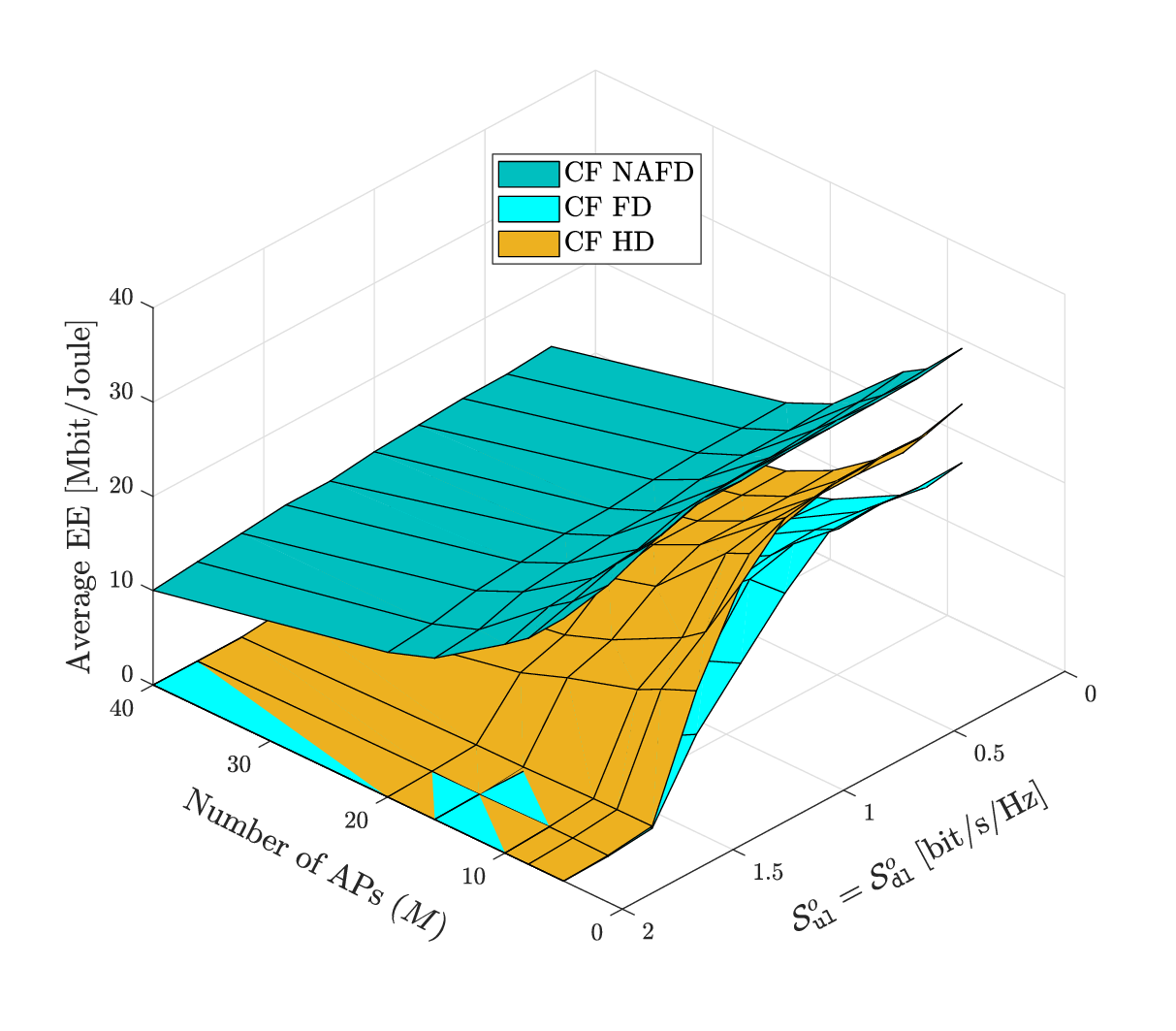} \vspace{-5mm}
    \caption{Average EE versus $\mathcal{S}_\dl^o= \mathcal{S}_\ul^o$ for different values of $M$  ($MN=80$, $K=L = 4$, $\tilde{\rho}_u=0.2$ Watt, $\tilde{\rho}_d=1$ Watt,  $\SIm/\Sn = 50$ dB, $B=50$ MHz).}
    \label{fig:nafd} \vspace{-1em}
\end{figure}

Figure~\ref{fig:nafd} compares the average EE performance of NAFD CF  and conventional CF system designs as a function of $\mathcal{S}\dl^o= \mathcal{S}\ul^o$ and $M$, with the number of service antennas kept fixed at $MN=40$. The path-loss model from~\cite{emil20TWC} has been considered and the parameters of the power consumption are set according to~\cite[Table II ]{Mohammadali:JSAC:2023}.  It can be observed that NAFD satisfactorily supports the SE requirements of both UL and DL UEs, while both the FD and HD setups fail to meet SE requirements greater than $1.2$ [bit/s/Hz].

\section{Challenges and Future Research Directions}~\label{sec:challenges}
This section addresses open challenges, trends, and the potential use of CF FD with other wireless technologies.
\subsection{mmWave Communications}
Due to its ample frequency spectrum resource, i.e., ranging from \qty{30}{\GHz} to \qty{100}{\GHz}, mmWave communication is a viable technology for next-generation wireless communication, offering a significantly higher capacity than existing wireless local area networks (WLANs) and cellular mobile communication \cite{Alkhateeb2014, Sun2014, Xiao2016}. However, a fundamental challenge of mmWave communication is the extremely high path loss caused by foliage, air absorption, rain-induced fading, and sensitivity to obstacles, resulting in lower received SINR. To overcome this issue and establish an adequate link budget, dense antenna arrays with dozens or hundreds of elements and joint transmit-receiver beamforming are required to enable substantial beamforming gains via directional transmissions.  Fortunately, owing to the small wavelength of the mmWave frequency, large antenna arrays can be packed into a compact area, allowing practical realization. On the other hand, in contrast to omnidirectional transmissions, which can interfere with many nodes, directional transmission is essential as it reduces the number of nodes affected by the interfering signal.

Nevertheless, recent developments in mmWave communication, including digital/analog beamforming with dense antenna arrays and a reduced number of RF chains, enable mmWave FD communication, overcoming significant path loss and establishing sufficient link margin \cite{Xiao2017, Satyanarayana2019, Roberts2022}. Although existing conventional FD (i.e., sub-\qty{6}{\GHz}) techniques cannot be directly applied to mmWave due to fundamental differences, such as propagation characteristics, dense antenna arrays, hybrid beamforming, wide bandwidths, high sampling rates, and beam alignment techniques, many novel transmission strategies have been developed for investigating the feasibility of mmWave FD communication \cite{Liangbin2014, Demir2016, Rajagopal2014, Xiao2017, Satyanarayana2019, Roberts2022, Barneto2022}. 

However, traditional cell-based mmWave FD will not be sufficient to meet the ever-increasing communication requirements \cite{Femenias2019}. Hence, integrating mmWave communications with the CF FD technology is a viable direction for next-generation wireless networks \cite{Femenias2019}. In particular, the macro-diversity gains achieved by distributed APs can compensate for the mmWave signal's limited coverage. Furthermore, mmWave CF FD communication will be able to realize a network with flexible duplex modes, efficiently underpin bidirectional communication utilizing CLI cancellation \cite{Wang2020}, and compensate for the orders-of-magnitude increase in free-space path loss caused by the use of higher frequencies with proper hybrid analog-digital beamforming \cite{Femenias2019}. Thus, this integration will push the limits of next-generation wireless networks. 

Only a few studies have looked into the viability of mmWave CF FD communication \cite{Fukue2023, Jiamin2023}. Reference \cite{Fukue2023} investigated a mmWave NAFD CF  system and proposed a cooperative resource allocation and beamforming design with location-aided channel estimation to minimize the inter-user interference. In particular, as mmWave channels are dominated by LoS paths, an inter-user channel estimate approach was proposed to reconstruct the channels among users using their locations. The joint AP mode selection and beamforming design were then developed using these channel estimates. Reference \cite{Jiamin2023} also studied a mmWave NAFD CF  system with DAC quantization and fronthaul compression. The authors looked into weighted UL and DL sum rate maximization by jointly optimizing the remote antenna unit's transmit power, UL user power allocation, and the DL and UL fronthaul compression noises variances. 

\subsection{UAV Communications}
UAVs, also known as drones, have caught the attention of the research community recently due to their autonomy, flexibility, and wide range of applications, including military, surveillance and monitoring, telecommunications, medical supply delivery, rescue operations, and more \cite{Li2019, Mozaffari2019, Wang2018, Song2019}. Although traditional UAV-centric research focuses on enhancing control and autonomy due to robotics or military applications, UAV-assisted communications, i.e., flying BSs or mobile relays, have recently emerged \cite{Li2019, Mozaffari2019, Wang2018, Song2019}. In such cases, UAVs serve as aerial communication platforms,   providing/improving communication services to ground targets in high traffic demand and overloaded situations. Due to the high altitude of UAVs, there is a greater likelihood of LoS links between the ground UEs and APs, resulting in improved channel quality compared to terrestrial networks. 

Recent UAV research has focused on integrating FD transceivers in UAV networks \cite{Wang2018, Song2019, Hua2020, Shi2021, Katwe2022, Liu2019, Zhu2020, Zhang2018Drone} and CF UAV communication \cite{Zheng2021, Zheng2021JSAC, Liu2021, Wu2023, Nahhas2022, Diaz2023, Andrea2020} to improve the SE.  While these approaches have been studied separately, combining UAVs with CF FD presents a promising yet unexplored research direction for enhancing UAV network performance. However, FD communication in low-altitude 3-D coverage scenarios is challenging due to increased UL and DL traffic and amplified interference from LoS links.  Recent works~\cite{Wan:TWC:2024,Wan:TCOM:2024,Li:IoT:2024} have introduced the NAFD concept into aerial-ground CF  systems to achieve flexible duplex communication in the spatial domain, allowing UAVs to select duplex modes based on the traffic distribution.

\subsection{Reconfigurable Intelligent  Surfaces}
RISs or intelligent reflecting surfaces provide cost-effective, energy-efficient, and spectrum-efficient communications~\cite{Wu2021, Diluka2020RIS, Galappaththige2022RIS,  Galappaththige2021Cellfree}. An RIS comprises a planar array of hundreds or thousands of ``nearly passive" reflecting elements. A software controller can control the reflective properties of these components individually, changing the real-time response to dynamic wireless channels, abrupt network changes, and/or traffic demands. In particular, the controller can customize individual elements to change their states to achieve coherent combining for the reflected signals at chosen receivers, thus increasing the received signal power. At the same time, the reflected signals combine destructively at non-intended receivers, suppressing co-channel interference or increasing the secrecy rate ~\cite{Wu2021, Diluka2020RIS}. Furthermore, because no transmit RF chains or amplifiers are required, an RIS panel may consume only a few Watts of power during the reconfiguration stages and significantly less during idle states. For $3$-, $4$-, $5$-, and $6$-bit resolution phase shifting, typical power consumption values per each phase shifter are respectively $1.5$, $4.5$, $6$, and $7.8$ mW~\cite{Huang2019}. Even without an amplifier, an RIS may produce significant gain—about $30$ to $40$ decibels relative to isotropic (dBi) depending on surface size and frequency~\cite{Poulakis2022}. As a result, an RIS has much lower hardware/energy costs than standard active antenna arrays~\cite{Hu2018, Wu2021, Liu2021}.

An RIS resembles an FD device without antenna noise amplification or SI, providing an efficient alternative for FD relays needing complicated SI cancellation methods. Although the integration of FD with RIS is being actively researched in traditional co-located MIMO setups~\cite{Vu2023, Efrem2023, Bazrafkan2023, Elhattab2022}, only a limited amount of work exists in HD CF systems \cite{Van:TWC:2022,Nguyen:TWC:2023,Thien:WCNC:2024} whilst the first papers on the CF FD with RIS are yet to appear.  

With the current investigations on RIS in many wireless systems, such as HD CF, mMIMO, IoT, and many others \cite{Vu2023, Efrem2023, Bazrafkan2023, Elhattab2022}, it can be anticipated that they will present significant advantages and opportunities for CF FD systems. By dynamically adjusting the phase and amplitude of reflected signals, RIS can enhance the signal strength, improve coverage, and facilitate better interference management. This capability is crucial for mitigating the SI inherent in FD operations. Furthermore, an RISs can bolster security by creating dynamic, adaptive signal paths that are harder for eavesdroppers to intercept. The improved SE and energy savings resulting from optimized signal reflections would also contribute to more robust and efficient CF FD networks.

\subsection{Integrated Sensing and Communications}
Among many visionary assumptions about $6$G  wireless networks, a common theme is that sensing will play a more significant role than ever before. It is anticipated that future wireless networks will be able to image and measure the surrounding environment, thereby enabling advanced location-aware services. This synergy between communication and sensing is referred to as integrated sensing and communications (ISAC) and  finds applications in various emerging areas, including vehicular networks, environmental monitoring, and human activity recognition. Recently, ISAC has garnered recognition from the International Telecommunication Union (ITU) as a pivotal element within the $6$G landscape, serving as one of six key $6$G usage scenarios~\cite{itu2030draft}. 

ISAC can be realized through a synergistic design of communications and radar/sensing systems that share spectral and hardware resources. This involves developing strategies for cohabitation designs and/or dynamic spectrum allocation and access, along with interference suppression or management. Alternatively, ISAC can be implemented from a co-design perspective, allowing communications and radar/sensing systems to operate in parallel with jointly optimized performance~\cite{Liu:JSAC:2022,Liu:TuTs:2022, Diluka2024NF, Azar2024}. Therefore, ISAC offers two primary advantages over dedicated sensing and communication functionalities. Firstly, it provides integration gains, enabling the efficient utilization of congested resources. Secondly, it offers coordination gains, allowing for the balancing of multi-functional performance and mutual assistance. The rapid evolution of ISAC suggests a future dual-functional network that could enable innovative services for civilians. In this network, the sensing operation collects and extracts information from noisy observations, while communication transfers information through specifically tailored signals and recovers it~\cite{Liu:JSAC:2022,Liu:TuTs:2022, Diluka2024NF, Azar2024}. 

ISAC has primarily been studied in single-cell scenarios with monostatic sensing, where the transmitter and receiver are co-located for sensing~\cite{Liu:TSP:2020,Liu:TWC:2020,Bazzi:TWC:2023, Diluka2023}. However, distributed designs can be employed to leverage macro diversity gain, thereby mitigating the effects of large-scale fading in single-cell scenarios. In line with this, the integration of CF  and ISAC has been explored in a few recent studies~\cite{Mao:TWC:2024,Behdad:TWC:2024,elfiatoure2024multiple}. The systems considered in these studies operate in a multi-static manner~\cite{Liu:TuTs:2022}, i.e., multiple non-co-located transmitters and receivers are deployed. Specifically, the design in~\cite{Mao:TWC:2024} assumed that all the APs serve the UEs over the same time-frequency resources, while the sensing signal for a specific target is generated by its designated AP (only some specific APs serve as an ISAC transmitter). The echo signals reflected by the target are received by all APs and sent back to the CPU for further processing (all APs serve as ISAC receivers). Consequently, the APs operate in FD mode to receive the reflected signals. However, from the sensing perspective, the main focus of this study is beampattern design at the transmitter, and the challenges related to the FD operation at the APs during the reception of echo signals remain unexplored. In~\cite{Behdad:TWC:2024}, an ISAC system with DL communication and multi-static sensing in a cloud radio access network (C-RAN) architecture has been considered, wherein each AP either serves as an ISAC transmitter or a sensing receiver. This architecture dispenses with the FD capability requirement in monostatic sensing while stimulating FD operation via HD APs and reflecting the use of the NAFD concept for ISAC CF  systems. However, the operation mode of the APs is fixed. A more flexible design has been proposed in~\cite{elfiatoure2024multiple}, where, depending on the network requirements, the operation mode of the APs is determined as either a communication or sensing AP. Inter-system interference is managed through joint operation mode selection and power control design at the APs.

\subsection{Machine Learning}
The application of ML in wireless communications has emerged as a transformative force, driving innovation and efficiency in the field. By leveraging sophisticated algorithms and data-driven insights, ML enhances various aspects of wireless communication systems, from optimizing signal processing and network management to improving spectrum utilization and security. This integration not only boosts the performance and reliability but also paves the way for the application of new advanced technologies in wireless networks. As wireless networks become increasingly complex, the role of ML in addressing challenges and unlocking new potentials continues to grow, marking a significant evolution in the future of wireless networks.

CF  networks have also availed of ML across multiple dimensions, including channel estimation~\cite{Athreya:WCL:2020}, power control~\cite{Bashar:JSAC:2020}, beamforming design~\cite{Eryani:JSAC:2021,Eryani:TCOM:2022}, and UE clustering within NOMA CF  systems~\cite{Eryani:JSAC:2021}. These studies leverage the capabilities of ML to address the challenges arising from high-dimensional and complex (nonlinear) optimization problems in the CF  context. ML methods have the ability to efficiently process and analyze vast datasets in real-time, making them ideal for large-scale deployments and effectively mitigating scalability concerns in these networks. Furthermore, ML techniques enable real-time decision-making for resource management, performance optimization, and low-latency communication. AP mode selection in NAFD CF  networks introduces a new dimension of complexity to the optimization problems. Consequently, ML can be leveraged to tackle these intricate optimization challenges~\cite{Sun:JSYS:2024,Wan:TWC:2024,Li:IoT:2024}.

\subsection{Unexplored Research Avenues}
Aside from the above-mentioned research directions, the integration of CF FD with other well-known technologies, including backscatter communications (\bc), mobile edge computing, rate splitting multiple access, vehicle-to-everything communication, underwater wireless communication,  the age of information, and more, remain untouched as of today.

\section{Conclusion}~\label{sec:conc}
This research investigated the potential and advantages of CF FD systems for the next generation of wireless networks. Initially, the fundamentals of FD communication and CF architecture were reviewed, highlighting their impact on current wireless systems. The integration of FD technology into CF architectures was then examined, focusing on channel estimation, performance analysis, and resource allocation. FD NAFD CF networks were also explored, supported by a comprehensive literature survey and case studies using a general CF FD system configuration. Additionally, emerging CF FD paradigms, such as mmWave communications, UAV communication, RIS, ISAC, and others, were discussed. Finally, several research gaps were identified, providing a foundation for future research efforts.

\balance

\bibliographystyle{IEEEtran}
\bibliography{IEEEabrv,Ref}

\end{document}

%% file: Figures/Tranceivers.eps_tex
\begingroup%
  \makeatletter%
  \providecommand\color[2][]{%
    \errmessage{(Inkscape) Color is used for the text in Inkscape, but the package 'color.sty' is not loaded}%
    \renewcommand\color[2][]{}%
  }%
  \providecommand\transparent[1]{%
    \errmessage{(Inkscape) Transparency is used (non-zero) for the text in Inkscape, but the package 'transparent.sty' is not loaded}%
    \renewcommand\transparent[1]{}%
  }%
  \providecommand\rotatebox[2]{#2}%
  \newcommand*\fsize{\dimexpr\f@size pt\relax}%
  \newcommand*\lineheight[1]{\fontsize{\fsize}{#1\fsize}\selectfont}%
  \ifx\svgwidth\undefined%
    \setlength{\unitlength}{552.68518066bp}%
    \ifx\svgscale\undefined%
      \relax%
    \else%
      \setlength{\unitlength}{\unitlength * \real{\svgscale}}%
    \fi%
  \else%
    \setlength{\unitlength}{\svgwidth}%
  \fi%
  \global\let\svgwidth\undefined%
  \global\let\svgscale\undefined%
  \makeatother%
  \begin{picture}(1,0.542824)%
    \lineheight{1}%
    \setlength\tabcolsep{0pt}%
    \put(0,0){\includegraphics[width=\unitlength]{Tranceivers.eps}}%
    \put(0.24362802,0.01610972){\color[rgb]{0,0,0}\makebox(0,0)[t]{\lineheight{1.25}\smash{\begin{tabular}[t]{c}(a) Seperate antenna transceiver\end{tabular}}}}%
    \put(0.23548595,0.38793079){\makebox(0,0)[t]{\lineheight{1.25}\smash{\begin{tabular}[t]{c}Scatterers/Reflectors\end{tabular}}}}%
    \put(0.75250723,0.01610972){\color[rgb]{0,0,0}\makebox(0,0)[t]{\lineheight{1.25}\smash{\begin{tabular}[t]{c}(b) Shared antenna transceiver\end{tabular}}}}%
    \put(0.74436516,0.38793079){\makebox(0,0)[t]{\lineheight{1.25}\smash{\begin{tabular}[t]{c}Scatterers/Reflectors\end{tabular}}}}%
    \put(0.11812748,0.11366502){\color[rgb]{0,0,0}\makebox(0,0)[t]{\lineheight{1.25}\smash{\begin{tabular}[t]{c}Transmitter\end{tabular}}}}%
    \put(0.1208415,0.08381077){\color[rgb]{0,0,0}\makebox(0,0)[t]{\lineheight{1.25}\smash{\begin{tabular}[t]{c}RF Chain\end{tabular}}}}%
    \put(0.25111456,0.29550453){\color[rgb]{0,0,0}\makebox(0,0)[t]{\lineheight{1.25}\smash{\begin{tabular}[t]{c}Direct SI\end{tabular}}}}%
    \put(0.25111456,0.20322777){\color[rgb]{0,0,0}\makebox(0,0)[t]{\lineheight{1.25}\smash{\begin{tabular}[t]{c}Leakage SI\end{tabular}}}}%
    \put(0.38952971,0.34164292){\color[rgb]{0,0,0}\makebox(0,0)[t]{\lineheight{1.25}\smash{\begin{tabular}[t]{c}Reflected SI\end{tabular}}}}%
    \put(0.84772243,0.26008243){\color[rgb]{0,0,0}\makebox(0,0)[t]{\lineheight{1.25}\smash{\begin{tabular}[t]{c}Direct SI\end{tabular}}}}%
    \put(0.85302384,0.34164292){\color[rgb]{0,0,0}\makebox(0,0)[t]{\lineheight{1.25}\smash{\begin{tabular}[t]{c}Reflected SI\end{tabular}}}}%
    \put(0.35967548,0.11366502){\color[rgb]{0,0,0}\makebox(0,0)[t]{\lineheight{1.25}\smash{\begin{tabular}[t]{c}Receiver\end{tabular}}}}%
    \put(0.3623895,0.08381077){\color[rgb]{0,0,0}\makebox(0,0)[t]{\lineheight{1.25}\smash{\begin{tabular}[t]{c}RF Chain\end{tabular}}}}%
    \put(0.62836371,0.11366502){\color[rgb]{0,0,0}\makebox(0,0)[t]{\lineheight{1.25}\smash{\begin{tabular}[t]{c}Transmitter\end{tabular}}}}%
    \put(0.63107773,0.08381077){\color[rgb]{0,0,0}\makebox(0,0)[t]{\lineheight{1.25}\smash{\begin{tabular}[t]{c}RF Chain\end{tabular}}}}%
    \put(0.86991171,0.11366502){\color[rgb]{0,0,0}\makebox(0,0)[t]{\lineheight{1.25}\smash{\begin{tabular}[t]{c}Receiver\end{tabular}}}}%
    \put(0.87262573,0.08381077){\color[rgb]{0,0,0}\makebox(0,0)[t]{\lineheight{1.25}\smash{\begin{tabular}[t]{c}RF Chain\end{tabular}}}}%
    \put(0.7599047,0.15929726){\color[rgb]{0,0,0}\makebox(0,0)[t]{\lineheight{1.25}\smash{\begin{tabular}[t]{c}Leakage SI\end{tabular}}}}%
  \end{picture}%
\endgroup%

%% file: Figures/FD_generic_block_diagram.eps_tex
\begingroup%
  \makeatletter%
  \providecommand\color[2][]{%
    \errmessage{(Inkscape) Color is used for the text in Inkscape, but the package 'color.sty' is not loaded}%
    \renewcommand\color[2][]{}%
  }%
  \providecommand\transparent[1]{%
    \errmessage{(Inkscape) Transparency is used (non-zero) for the text in Inkscape, but the package 'transparent.sty' is not loaded}%
    \renewcommand\transparent[1]{}%
  }%
  \providecommand\rotatebox[2]{#2}%
  \newcommand*\fsize{\dimexpr\f@size pt\relax}%
  \newcommand*\lineheight[1]{\fontsize{\fsize}{#1\fsize}\selectfont}%
  \ifx\svgwidth\undefined%
    \setlength{\unitlength}{203.33338219bp}%
    \ifx\svgscale\undefined%
      \relax%
    \else%
      \setlength{\unitlength}{\unitlength * \real{\svgscale}}%
    \fi%
  \else%
    \setlength{\unitlength}{\svgwidth}%
  \fi%
  \global\let\svgwidth\undefined%
  \global\let\svgscale\undefined%
  \makeatother%
  \begin{picture}(1,1.41919015)%
    \lineheight{1}%
    \setlength\tabcolsep{0pt}%
    \put(0,0){\includegraphics[width=\unitlength]{FD_generic_block_diagram.eps}}%
    \put(-1.83367189,3.26661911){\color[rgb]{0,0,0}\makebox(0,0)[lt]{\begin{minipage}{69.07594891\unitlength}\raggedright \end{minipage}}}%
    \put(0.07429171,0.11100947){\color[rgb]{0,0,0}\rotatebox{0.53733523}{\makebox(0,0)[lt]{\lineheight{1.25}\smash{\begin{tabular}[t]{l}Signal Processing\end{tabular}}}}}%
    \put(0.11855393,0.16264857){\color[rgb]{0,0,0}\rotatebox{0.53733522}{\makebox(0,0)[lt]{\lineheight{1.25}\smash{\begin{tabular}[t]{l}Transmitter\end{tabular}}}}}%
    \put(0.42197926,1.28214431){\color[rgb]{0,0,0}\rotatebox{0.53733522}{\makebox(0,0)[lt]{\lineheight{1.25}\smash{\begin{tabular}[t]{l}\textbf{Antenna}\end{tabular}}}}}%
    \put(0.39150519,1.24174606){\color[rgb]{0,0,0}\rotatebox{0.53733522}{\makebox(0,0)[lt]{\lineheight{1.25}\smash{\begin{tabular}[t]{l}\textbf{Cancellation}\end{tabular}}}}}%
    \put(0.64232372,0.11026934){\color[rgb]{0,0,0}\rotatebox{0.53733523}{\makebox(0,0)[lt]{\lineheight{1.25}\smash{\begin{tabular}[t]{l}Signal Processing\end{tabular}}}}}%
    \put(0.70871704,0.16190826){\color[rgb]{0,0,0}\rotatebox{0.53733523}{\makebox(0,0)[lt]{\lineheight{1.25}\smash{\begin{tabular}[t]{l}Receiver\end{tabular}}}}}%
    \put(0.74560227,1.13108964){\color[rgb]{0,0,0}\rotatebox{0.53733522}{\makebox(0,0)[lt]{\lineheight{1.25}\smash{\begin{tabular}[t]{l}LNA\end{tabular}}}}}%
    \put(0.68658596,0.96141772){\color[rgb]{0,0,0}\rotatebox{0.53733522}{\makebox(0,0)[lt]{\lineheight{1.25}\smash{\begin{tabular}[t]{l}\textbf{Analog/RF}\end{tabular}}}}}%
    \put(0.67920887,0.90977889){\color[rgb]{0,0,0}\rotatebox{0.53733522}{\makebox(0,0)[lt]{\lineheight{1.25}\smash{\begin{tabular}[t]{l}\textbf{Cancellation}\end{tabular}}}}}%
    \put(0.66445481,0.77699189){\color[rgb]{0,0,0}\rotatebox{0.53733522}{\makebox(0,0)[lt]{\lineheight{1.25}\smash{\begin{tabular}[t]{l}Downconverter\end{tabular}}}}}%
    \put(0.64232372,0.72535287){\color[rgb]{0,0,0}\rotatebox{0.53733523}{\makebox(0,0)[lt]{\lineheight{1.25}\smash{\begin{tabular}[t]{l}(RF to Baseband)\end{tabular}}}}}%
    \put(0.7456023,0.53355022){\color[rgb]{0,0,0}\rotatebox{0.53733522}{\makebox(0,0)[lt]{\lineheight{1.25}\smash{\begin{tabular}[t]{l}ADC\end{tabular}}}}}%
    \put(0.72347117,0.35650048){\color[rgb]{0,0,0}\rotatebox{0.53733522}{\makebox(0,0)[lt]{\lineheight{1.25}\smash{\begin{tabular}[t]{l}\textbf{Digital}\end{tabular}}}}}%
    \put(0.67183188,0.30486168){\color[rgb]{0,0,0}\rotatebox{0.53733522}{\makebox(0,0)[lt]{\lineheight{1.25}\smash{\begin{tabular}[t]{l}\textbf{Cancellation}\end{tabular}}}}}%
    \put(0.19232464,1.13109025){\color[rgb]{0,0,0}\rotatebox{0.53733522}{\makebox(0,0)[lt]{\lineheight{1.25}\smash{\begin{tabular}[t]{l}PA\end{tabular}}}}}%
    \put(0.07429172,0.71797575){\color[rgb]{0,0,0}\rotatebox{0.53733523}{\makebox(0,0)[lt]{\lineheight{1.25}\smash{\begin{tabular}[t]{l}(Baseband to RF)\end{tabular}}}}}%
    \put(0.11117687,0.76961482){\color[rgb]{0,0,0}\rotatebox{0.53733523}{\makebox(0,0)[lt]{\lineheight{1.25}\smash{\begin{tabular}[t]{l}Upconverter\end{tabular}}}}}%
    \put(0.1674616,0.53628262){\color[rgb]{0,0,0}\rotatebox{0.53733522}{\makebox(0,0)[lt]{\lineheight{1.25}\smash{\begin{tabular}[t]{l}DAC\end{tabular}}}}}%
    \put(0.10347433,0.32540415){\color[rgb]{0,0,0}\rotatebox{0.53733522}{\makebox(0,0)[lt]{\lineheight{1.25}\smash{\begin{tabular}[t]{l}$x_b(n)$\end{tabular}}}}}%
    \put(0.03576077,0.01127721){\color[rgb]{0,0,0}\rotatebox{0.53733522}{\makebox(0,0)[lt]{\lineheight{1.25}\smash{\begin{tabular}[t]{l}Transmitter RF Chain\end{tabular}}}}}%
    \put(0.62592472,0.01127725){\color[rgb]{0,0,0}\rotatebox{0.53733522}{\makebox(0,0)[lt]{\lineheight{1.25}\smash{\begin{tabular}[t]{l}Receiver RF Chain\end{tabular}}}}}%
    \put(0.1108517,0.62786522){\color[rgb]{0,0,0}\rotatebox{0.53733522}{\makebox(0,0)[lt]{\lineheight{1.25}\smash{\begin{tabular}[t]{l}$x_b(t)$\end{tabular}}}}}%
    \put(0.10347505,0.9303243){\color[rgb]{0,0,0}\rotatebox{0.53733522}{\makebox(0,0)[lt]{\lineheight{1.25}\smash{\begin{tabular}[t]{l}$x_u(t)$\end{tabular}}}}}%
    \put(0.11085245,1.2991766){\color[rgb]{0,0,0}\rotatebox{0.53733522}{\makebox(0,0)[lt]{\lineheight{1.25}\smash{\begin{tabular}[t]{l}$x_t(t)$\end{tabular}}}}}%
    \put(0.78953959,1.29917633){\color[rgb]{0,0,0}\rotatebox{0.53733522}{\makebox(0,0)[lt]{\lineheight{1.25}\smash{\begin{tabular}[t]{l}$y_r(t)$\end{tabular}}}}}%
  \end{picture}%
\endgroup%

%% file: Figures/SystemFigure.eps_tex
\begingroup%
  \makeatletter%
  \providecommand\color[2][]{%
    \errmessage{(Inkscape) Color is used for the text in Inkscape, but the package 'color.sty' is not loaded}%
    \renewcommand\color[2][]{}%
  }%
  \providecommand\transparent[1]{%
    \errmessage{(Inkscape) Transparency is used (non-zero) for the text in Inkscape, but the package 'transparent.sty' is not loaded}%
    \renewcommand\transparent[1]{}%
  }%
  \providecommand\rotatebox[2]{#2}%
  \newcommand*\fsize{\dimexpr\f@size pt\relax}%
  \newcommand*\lineheight[1]{\fontsize{\fsize}{#1\fsize}\selectfont}%
  \ifx\svgwidth\undefined%
    \setlength{\unitlength}{604.50915527bp}%
    \ifx\svgscale\undefined%
      \relax%
    \else%
      \setlength{\unitlength}{\unitlength * \real{\svgscale}}%
    \fi%
  \else%
    \setlength{\unitlength}{\svgwidth}%
  \fi%
  \global\let\svgwidth\undefined%
  \global\let\svgscale\undefined%
  \makeatother%
  \begin{picture}(1,0.56991801)%
    \lineheight{1}%
    \setlength\tabcolsep{0pt}%
    \put(0,0){\includegraphics[width=\unitlength]{SystemFigure.eps}}%
    \put(0.34307272,0.13049339){\makebox(0,0)[t]{\lineheight{1.25}\smash{\begin{tabular}[t]{c}${\rm{AP}}_m$\end{tabular}}}}%
    \put(0.70312933,0.3835913){\makebox(0,0)[t]{\lineheight{1.25}\smash{\begin{tabular}[t]{c}${\rm{U}}_k^{\mathtt{dl}}$\end{tabular}}}}%
    \put(0.70066032,0.06101554){\makebox(0,0)[t]{\lineheight{1.25}\smash{\begin{tabular}[t]{c}${\rm{U}}_l^{\mathtt{ul}}$\end{tabular}}}}%
    \put(0.91521049,0.31659479){\makebox(0,0)[t]{\lineheight{1.25}\smash{\begin{tabular}[t]{c}${\rm{U}}_1^{\mathtt{dl}}$\end{tabular}}}}%
    \put(0.88923073,0.47788267){\makebox(0,0)[t]{\lineheight{1.25}\smash{\begin{tabular}[t]{c}${\rm{U}}_K^{\mathtt{dl}}$\end{tabular}}}}%
    \put(0.91769184,0.12304934){\makebox(0,0)[t]{\lineheight{1.25}\smash{\begin{tabular}[t]{c}${\rm{U}}_1^{\mathtt{ul}}$\end{tabular}}}}%
    \put(0.90160024,0.02255458){\makebox(0,0)[t]{\lineheight{1.25}\smash{\begin{tabular}[t]{c}${\rm{U}}_L^{\mathtt{ul}}$\end{tabular}}}}%
    \put(0.1941916,0.04364607){\makebox(0,0)[t]{\lineheight{1.25}\smash{\begin{tabular}[t]{c}${\rm{AP}}_M$\end{tabular}}}}%
    \put(0.05771724,0.20493395){\makebox(0,0)[t]{\lineheight{1.25}\smash{\begin{tabular}[t]{c}${\rm{AP}}_1$\end{tabular}}}}%
    \put(0.49134354,0.43640768){\makebox(0,0)[t]{\lineheight{1.25}\smash{\begin{tabular}[t]{c}$\mathbf{f}_{mk}$\end{tabular}}}}%
    \put(0.34459993,0.42347944){\makebox(0,0)[t]{\lineheight{1.25}\smash{\begin{tabular}[t]{c}$\mathbf{Q}_{mm}$\end{tabular}}}}%
    \put(0.19661088,0.40181585){\makebox(0,0)[t]{\lineheight{1.25}\smash{\begin{tabular}[t]{c}$\mathbf{Q}_{m1}$\end{tabular}}}}%
    \put(0.55811654,0.37056114){\makebox(0,0)[t]{\lineheight{1.25}\smash{\begin{tabular}[t]{c}$\bar{\mathbf{f}}_{mk}$\end{tabular}}}}%
    \put(0.49587072,0.21457596){\makebox(0,0)[t]{\lineheight{1.25}\smash{\begin{tabular}[t]{c}$\mathbf{g}_{ml}$\end{tabular}}}}%
    \put(0.72820471,0.27668045){\makebox(0,0)[t]{\lineheight{1.25}\smash{\begin{tabular}[t]{c}$g_{kl}$\end{tabular}}}}%
    \put(0.5605979,0.27444169){\makebox(0,0)[t]{\lineheight{1.25}\smash{\begin{tabular}[t]{c}$\bar{\mathbf{g}}_{ml}$\end{tabular}}}}%
    \put(0.48914707,0.01580443){\makebox(0,0)[t]{\lineheight{1.25}\smash{\begin{tabular}[t]{c}CPU\end{tabular}}}}%
  \end{picture}%
\endgroup%

%% file: Figures/SystemFigure_NACF.eps_tex
\begingroup%
  \makeatletter%
  \providecommand\color[2][]{%
    \errmessage{(Inkscape) Color is used for the text in Inkscape, but the package 'color.sty' is not loaded}%
    \renewcommand\color[2][]{}%
  }%
  \providecommand\transparent[1]{%
    \errmessage{(Inkscape) Transparency is used (non-zero) for the text in Inkscape, but the package 'transparent.sty' is not loaded}%
    \renewcommand\transparent[1]{}%
  }%
  \providecommand\rotatebox[2]{#2}%
  \newcommand*\fsize{\dimexpr\f@size pt\relax}%
  \newcommand*\lineheight[1]{\fontsize{\fsize}{#1\fsize}\selectfont}%
  \ifx\svgwidth\undefined%
    \setlength{\unitlength}{604.50915527bp}%
    \ifx\svgscale\undefined%
      \relax%
    \else%
      \setlength{\unitlength}{\unitlength * \real{\svgscale}}%
    \fi%
  \else%
    \setlength{\unitlength}{\svgwidth}%
  \fi%
  \global\let\svgwidth\undefined%
  \global\let\svgscale\undefined%
  \makeatother%
  \begin{picture}(1,0.56991801)%
    \lineheight{1}%
    \setlength\tabcolsep{0pt}%
    \put(0,0){\includegraphics[width=\unitlength]{SystemFigure_NACF.eps}}%
    \put(0.63972475,0.5318997){\makebox(0,0)[t]{\lineheight{1.25}\smash{\begin{tabular}[t]{c}${\rm{U}}_k^{\mathtt{dl}}$\end{tabular}}}}%
    \put(0.11637134,0.26313204){\makebox(0,0)[t]{\lineheight{1.25}\smash{\begin{tabular}[t]{c}${\rm{U}}_1^{\mathtt{ul}}$\end{tabular}}}}%
    \put(0.87768649,0.24822104){\makebox(0,0)[t]{\lineheight{1.25}\smash{\begin{tabular}[t]{c}${\rm{U}}_1^{\mathtt{dl}}$\end{tabular}}}}%
    \put(0.90133377,0.46906137){\makebox(0,0)[t]{\lineheight{1.25}\smash{\begin{tabular}[t]{c}${\rm{U}}_{K_d}^{\mathtt{dl}}$\end{tabular}}}}%
    \put(0.32959639,0.53004634){\makebox(0,0)[t]{\lineheight{1.25}\smash{\begin{tabular}[t]{c}${\rm{U}}_l^{\mathtt{ul}}$\end{tabular}}}}%
    \put(0.07957043,0.47060477){\makebox(0,0)[t]{\lineheight{1.25}\smash{\begin{tabular}[t]{c}${\rm{U}}_{K_u}^{\mathtt{ul}}$\end{tabular}}}}%
    \put(0.71223448,0.11058145){\makebox(0,0)[t]{\lineheight{1.25}\smash{\begin{tabular}[t]{c}HD DL ${\rm{AP}}_M$\end{tabular}}}}%
    \put(0.24333154,0.11395488){\makebox(0,0)[t]{\lineheight{1.25}\smash{\begin{tabular}[t]{c}HD UL ${\rm{AP}}_1$\end{tabular}}}}%
    \put(0.47940947,0.00683423){\makebox(0,0)[t]{\lineheight{1.25}\smash{\begin{tabular}[t]{c}FD ${\rm{AP}}_m$\end{tabular}}}}%
    \put(0.91778104,0.01473089){\makebox(0,0)[t]{\lineheight{1.25}\smash{\begin{tabular}[t]{c}CPU\end{tabular}}}}%
    \put(0.42959378,0.30350212){\makebox(0,0)[lt]{\lineheight{1.25}\smash{\begin{tabular}[t]{l}$\mathbf{Q}_{mm}$\end{tabular}}}}%
    \put(0.5912852,0.25036624){\makebox(0,0)[lt]{\lineheight{1.25}\smash{\begin{tabular}[t]{l}$\mathbf{Q}_{mM}$\end{tabular}}}}%
    \put(0.44702439,0.5049824){\makebox(0,0)[lt]{\lineheight{1.25}\smash{\begin{tabular}[t]{l}$h_{kl}$\end{tabular}}}}%
    \put(0.60356065,0.36922845){\makebox(0,0)[lt]{\lineheight{1.25}\smash{\begin{tabular}[t]{l}$\mathbf{f}_{Mk}^{\mathtt{dl}}$\end{tabular}}}}%
    \put(0.28346625,0.36922845){\makebox(0,0)[lt]{\lineheight{1.25}\smash{\begin{tabular}[t]{l}$\mathbf{g}_{1l}^{\mathtt{ul}}$\end{tabular}}}}%
  \end{picture}%
\endgroup%